\documentclass[11pt, a4paper]{article}
\usepackage[top=0.8in, bottom=0.8in, left=1.0in, right=1.0in]{geometry}
\usepackage{booktabs}
\usepackage{multirow}
\usepackage{setspace}
\usepackage{natbib}
\usepackage{authblk}
\usepackage[pdftex]{graphicx}
\usepackage{amsmath}
\usepackage{amsthm}
\usepackage{mathtools}
\usepackage{amsfonts}
\usepackage{ascmac}
\usepackage{amssymb}
\usepackage{bm}
\usepackage{color}
\usepackage{enumitem}
\usepackage{mathrsfs}
\usepackage[ 
	colorlinks = true, 
	anchorcolor=black,
	linkcolor = blue,
	citecolor = blue, 
	filecolor  = black, 
	urlcolor   = blue
]{hyperref}
\usepackage{comment}
\usepackage{algorithmic, algorithm}

\usepackage{breqn}
\usepackage[subrefformat=parens]{subcaption}
\allowdisplaybreaks[1]

\newcommand{\biblist}{\begin{list}{}
{\listparindent 0.0cm \leftmargin 0.50cm \itemindent -0.50 cm
\labelwidth 0 cm \labelsep 0.50 cm
\usecounter{list}}\clubpenalty4000\widowpenalty4000}
\newcommand{\ebiblist}{\end{list}}

 \theoremstyle{plain}
\newtheorem{thm}{Theorem}

\theoremstyle{definition}

\newtheorem{exm}{Example}

\theoremstyle{remark}
\newtheorem*{rem}{Remark}

\newcommand{\ep}{\varepsilon}

\title{\bf Robustness of Bayesian ordinal response model against outliers via divergence approach}
\author[1]{Tomotaka Momozaki}
\author[2]{Tomoyuki Nakagawa}
\affil[1]{Department of Information Sciences, Tokyo University of Science}
\affil[2]{School of Data Science, Meisei University}

\date{Last update: \today}

\begin{document}
\maketitle
%=abstract=========================================================
\begin{abstract}
Ordinal response model is a popular and commonly used regression for ordered categorical data in a wide range of fields such as medicine and social sciences.
However, it is empirically known that the existence of ``outliers'', combinations of the ordered categorical response and covariates that are heterogeneous compared to other pairs, makes the inference with the ordinal response model unreliable.
In this article, we prove that the posterior distribution in the ordinal response model does not satisfy the posterior robustness with any link functions, i.e., the posterior cannot ignore the influence of large outliers.
Furthermore, to achieve robust Bayesian inference in the ordinal response model, this article defines general posteriors in the ordinal response model with two robust divergences (the density-power and $\gamma$-divergences) based on the framework of the general posterior inference.
We also provide an algorithm for generating posterior samples from the proposed posteriors.
The robustness of the proposed methods against outliers is clarified from the posterior robustness and the index of robustness based on the Fisher-Rao metric.
Through numerical experiments on artificial data and two real datasets, we show that the proposed methods perform better than the ordinary bayesian methods with and without outliers in the data for various link functions.
\end{abstract}


\medskip

%=introduction=========================================================
\section{Introduction}
\label{sec:intro}
Ordered categorical data appear frequently in a wide range of fields such as medicine, sociology, psychology, political sciences, economics, marketing, and so on, and studies on their analysis methods have been quite active even today (\citealp{franses2001quantitative, rossi2003bayesian, agresti2010analysis, madahian2015bayesian, agresti2017ordinal, satake2018sparse, liu2018residuals, baetschmann2020feologit}).
Ordered categorical data are, for example, the progression of a certain disease expressed as stage 1, 2, 3, or 4, or opinions on a certain policy expressed as opposition, neutrality, or approval.
Additionally, when continuous data are summarized into categorical data, such as ages 0-20, 21-40, 41-60, 61-80, and 80 or more, the categorical data are ordered categorical data.
For this reason, ordered categorical data are often considered to be discretized values of latent continuous variables.
For more details on ordered categorical data, its traditional analysis methods, and related studies, we refer the reader to \cite{agresti2010analysis}, which is a very excellent book.

The ordinal response model has been gaining popularity in regression for the ordered categorical data thanks to its interpretability and flexibility since it was proposed by a pioneering work \cite{walker1967estimation} in this field.
The ordinal response model is also being developed as a package in {\bf R} (\citealp{r2022r}), one of the most popular programming languages today, and the ease that the analysis can be performed regardless of the maximum likelihood or Bayesian methods is another reason why the ordinal response model is often used.
For details on the {\bf R} packages for the ordinal response model, see \cite{christensen2018cumulative}.
The ordinal response model is one of the frameworks of generalized linear models, and is called the ordinal regression model, or the cumulative link model since it connects the cumulative probability of belonging to a certain category to the covariates with the ``link'' function (\citealp{mccullagh1980regression}). 
The link functions most commonly used the probit, logit, log-log, and complementary log-log (clog-log) links, which are the distribution function of the standard normal, logistic, right-skewed Gumbel, and left-skewed log-Weibull distributions, respectively.
The formulation of the ordinal response model is described in Section \ref{sec:borm}.
The inference methods in the ordinal response model are generally the maximum likelihood method (\citealp{mccullagh1980regression, harrell2001regression}) and the Bayesian method with latent variables (\citealp{albert1993bayesian, nandram1996reparameterizing}), but these methods are known to be strongly affected by outliers in the data.
An ``outlier'' in the ordinal response model is defined as a combination of ordered categorical data and its covariates that is heterogeneous relative to other pairs (\citealp{riani2011outliers}).
This may be taken to mean that the combination of ordered categorical data and its covariates is inconsistent.
Outliers can be caused by various reasons, such as typos in the values and misrecognition of units.
One of the other common examples of outliers appearing is that in social surveys such as questionnaires, respondents sometimes submit incoherent answers due to their lack of interest in the issue.

There are various ways to check whether an inference method is robust against outliers in data, and inference in the frequentist framework may often use the influence function mentioned in \cite{hampel1974influence}.
Simply put, the influence function is an index to check the ``influence'' of the data on an inference method, and its value must be at least bounded to be robust against outliers.
This is because if the values diverge for certain data, the inference results will be almost entirely dependent on that data, ignoring other data.
The influence function is desired to satisfy not only boundedness but also redescendence. 
While boundedness means that the influence from outliers is limited but a certain amount of influence still remains, redescendence means that the influence of large outliers on the inference can be ignored (\citealp{maronna2019robust}).
Hence, many robust inference methods based on the Huber-loss and robust divergences (\citealp{hampel1986robust, basu1998robust, jones2001comparison, fujisawa2008robust, huber2009robust, ghosh2013robust, maronna2019robust, castilla2021estimation}) have been studied so that the influence function satisfies boundedness and redescendence.
However, most of them are focused on continuous, binary, or counted data.

In some studies on robust inference methods in the ordinal response model, \cite{croux2013robust} and \cite{iannario2017robust} proposed the weighted maximum likelihood methods using the Student, 0/1, and Huber weights so that the influence function is bounded.
\cite{scalera2021robust} derived a class of link functions for bounded influence functions in the maximum likelihood method in the ordinal response model, and \cite{momozaki2022robustness} proved that the maximum likelihood method in the ordinal response model can no longer satisfy the redescendence of the influence function.
Their contributions show that the maximum likelihood method in the ordinal response model does not allow robust and flexible analysis (the conditions derived by \cite{scalera2021robust} are not satisfied by commonly used link functions such as the probit and logit links), and that more robust inference in the maximum likelihood method is no longer achievable.
Unfortunately, \cite{czado1992effect} showed that misspecification of the link function in binary regression causes a substantial bias in the parameter inferences, and it is confirmed that the same thing happens in the ordinal response model (\citealp{momozaki2022robustness}), so the inability to select the link function is very serious problem.

One way to solve these problems is to develop inference methods using robust divergences such as the density-power and $\gamma$-divergences (\citealp{basu1998robust, jones2001comparison, fujisawa2008robust}), which have contributed to the development of robust inference methods in the framework of linear regression and others.
\cite{pyne2022robust} and \cite{momozaki2022robustness} proposed the inference methods in the ordinal response model using the density-power and $\gamma$-divergences.
\cite{momozaki2022robustness} further derived conditions for the link function to satisfy both boundedness and redescendence of the influence functions in those inference methods.
Since these conditions are satisfied by commonly used link functions such as the probit, logit, log-log, and clog-log link functions, their contributions allow analysts to perform robust and flexible analysis.
Thus, in the framework of the frequentist approach, there is a certain amount of research on robust inference methods in the ordinal response model, although the number of such methods is limited.
However, to the best of our knowledge, there are no studies on robust Bayesian inference methods in the ordinal response model, although there have been many studies on robust Bayesian inference on continuous, binary, or counted data.

In the Bayesian framework, there are methods to check whether the Bayesian inference method is robust against outliers in the data, such as the Bayesian version of the influence function mentioned in \cite{ghosh2016robust} and \cite{nakagawa2019robust}, or the index using the Fisher-Rao metric proposed in \cite{kurtek2015bayesian}.
Another method is to prove that the posterior robustness (\citealp{desgagne2015robustness}), which is the property that the influence of outliers on the posterior distribution of parameters in a statistical model is negligible even if there are large outliers in the data, holds.
There are also many studies on the Bayesian inference methods using the Huber-loss and robust divergences, for example, \cite{kawakami2023approximate} proposed the linear model using the Huber-loss, \cite{ghosh2016robust} proposed the inference method of mean parameter using the density-power divergence, \cite{nakagawa2019robust} proposed the inference method of mean and variance parameters using the $\gamma$-divergence, and \cite{hashimoto2020robust} proposed the inference method in the linear model using the $\gamma$-divergence.
\cite{jewson2018robust} provides a comprehensive review in Bayesian inferences with the robust divergences.
Another robust Bayesian inference methods are those using distributions with the extremely heavy tail, for example, the log-Pareto truncated normal distribution proposed by \cite{gagnon2020new} and the extremely heavily-tailed distribution proposed by \cite{hamura2022log}.

In this article, we prove that the posterior distribution of parameters based on the log-likelihood in the ordinal response model does not satisfy the posterior robustness with any link functions.
Furthermore, to achieve robust Bayesian inference in the ordinal response model, we define general (synthetic) posteriors of the parameters based on the framework of the general (synthetic) posterior inference (\citealp{bissiri2016general, jewson2018robust, bhattacharya2019bayesian, miller2018robust, nakagawa2019robust, hashimoto2020robust}), by replacing the log-likelihood function with two robust divergences, the density-power and $\gamma$-divergences.
The robustness of these proposed methods against outliers is clarified from the posterior robustness and the index of robustness proposed by \cite{kurtek2015bayesian}.
Although sampling from the proposed posteriors is apparently intractable, we solve this problem by using the Bayesian bootstrap (\citealp{rubin1981bayesian, newton1994approximate, newton2021weighted}).
Numerical experiments using several artificial data with varying outlier ratios show that the proposed methods achieve better estimation accuracy than conventional one, regardless of the presence or absence of outliers.
Two real data applications are also used to demonstrate the usefulness of the proposed methods.

This article is organized as follows.
Section \ref{sec:borm} introduces the ordinal response model and the posterior distribution based on the log-likelihood.
Section \ref{sec:rborm} defines Bayesian ordinal response models using the density-power and $\gamma$-divergences, and formulates the posteriors in those models.
We also describe the sampling method from the proposed posteriors using the WLB method.
Section \ref{sec:post_robust} proves that posterior based on the log-likelihood in the ordinal response model does not satisfy the posterior robustness, and that the proposed posteriors in our robust Bayesian ordinal response models satisfies the posterior robustness.
Section \ref{sec:ne} confirms that the proposed Bayesian ordinal response models with the robust divergences are robust against outliers by numerical experiments using the index of robustness proposed by \cite{kurtek2015bayesian} and by applying them to artificial data and some real data.
Section \ref{sec:conclude} provides the conclusion and some remarks.

%=sec:borm=========================================================
\section{Bayesian ordinal response model and its problems}
\label{sec:borm}
In this section, we discuss the ordinal response model, several Bayesian inference methods in this model, and problems with those methods in the presence of outliers in data.

The probability mass function for an $i$-th unit ($i=1,2,\ldots,n$) in the ordinal response model can be expressed as 
\begin{equation*}
f(y_i|\bm{x}_i;\bm{\theta}) = \Pr(y_i=m|\bm{x}_i;\bm{\theta}) = G(\delta_{y_i}-\bm{x}_i^\top\bm{\beta}) - G(\delta_{y_i-1}-\bm{x}_i^\top\bm{\beta})
\end{equation*}
by considering the following latent variable model for $M$ ordered categorical data $y_i$ as the response variable.
\begin{equation}
\label{eq:lvm}
z_i = \bm{x}_i^\top \bm{\beta} + \ep_i,
\end{equation}
for $i=1,2,\ldots,n$, where $z_i$ is called the (continuous) latent variable, $y_i=m \Leftrightarrow \delta_{m-1}<z_i\leq\delta_m$ ($m=1,2,\ldots,M$), the covariates $\bm{x}_i = (x_{i1}, x_{i2}, \ldots, x_{ip})^\top$, the coefficients $\bm{\beta} = (\beta_1, \beta_2, \ldots, \beta_p)^\top$, $\ep_i \overset{i.i.d.}{\sim} G(\cdot)$ (known) and has the density function $g(\cdot)$, the cutpoints $-\infty=\delta_0<\delta_1<\cdots<\delta_M=\infty$, and the parameters for inference $\bm{\theta}=(\bm{\beta},\bm{\delta})$ with $\bm{\delta}=(\delta_1,\delta_2,\ldots,\delta_{M-1})$.
We refer to $G(\cdot)$ as the link function.
For the identification, in the $G(\cdot)$ there is no error scale and the latent model has no the intercept term.
Note that the observed data is $(\bm{y}, \bm{X})$, where $\bm{y}=(y_1,y_2,\ldots,y_n)^\top$ and $\bm{X}=(\bm{x}_1,\bm{x}_2,\ldots,\bm{x}_n)^\top$, and the latent variable $\bm{z}=(z_1,z_2,\ldots,z_p)^\top$ is unobserved.
The likelihood function in the ordinal response model is 
\begin{equation}
\label{eq:orm}
f(\bm{y}|\bm{X};\bm{\theta}) = \prod_{i=1}^n \left[ G(\delta_{y_i}-\bm{x}_i^\top\bm{\beta}) - G(\delta_{y_i-1}-\bm{x}_i^\top\bm{\beta}) \right].
\end{equation}
Let $p(\bm{\theta})$ be a prior distribution for the parameters $\bm{\theta}$.
Then the (standard) posterior distribution of $\bm{\theta}$ is given by
\begin{equation}
\label{eq:post}
p(\bm{\theta}|D) \propto p(\bm{\theta}) f(\bm{y}|\bm{X};\bm{\theta}) = p(\bm{\theta}) \exp\left\{ \sum_{i=1}^n \log \left[ G(\delta_{y_i}-\bm{x}_i^\top\bm{\beta}) - G(\delta_{y_i-1}-\bm{x}_i^\top\bm{\beta}) \right] \right\},
\end{equation}
where $D$ denotes the set of observed data.

The posterior samplings of $\bm{\theta}$ from equation \eqref{eq:post} are generally intractable.
\cite{albert1993bayesian} addressed this problem in the probit and robit ($G(\cdot)$ is the distribution function of the Student's $t$ distribution) links by using the data augmentation method with latent variables $\bm{z}$.
In the case of the probit link, assume the flat prior for $\bm{\theta}$ ($p(\bm{\theta}) \propto c$), the full conditional posterior of $\bm{\beta}$, $\delta_j$ ($j=1,2,\ldots,M-1$), and $z_i$ ($i=1,2,\ldots,n$) are the multivariate normal, the uniform, and the  truncated normal distributions, respectively, so the posterior samplings from these distributions are tractable.
Note that \cite{albert1993bayesian}'s method is known to reduce the effective sample size due to the high autocorrelation of Gibbs samplings of the cutpoints parameter $\bm{\delta}$.
For the details to address this problem, see \cite{nandram1996reparameterizing} and \cite{sha2019bayes}.
Other methods for the posterior sampling of $\bm{\theta}$ include the Hamiltonian Monte Carlo method (\citealp{duane1987hybrid, neal2003slice, neal2011mcmc}) and its extension the no-U-turn sampler (\citealp{hoffman2014no}), and in the {\bf R} programming language, the \texttt{brm} and \texttt{stan\_polr} functions in the {\bf brms} (\citealp{burkner2017brms}) and {\bf rstanarm} (\citealp{rstanarm2020}) packages, respectively, using the probabilistic programming language {\bf Stan} (\citealp{carpenter2017stan}) are available for implementing them.

These Bayesian inferences in the ordinal response model are very useful for the analysis of ordered categorical data by virtue of their ease of implementation.
However, when there are outliers in the data, i.e., when there are heterogeneous pairs of ordered categorical data and covariates $(\bm{y}, \bm{X})$ compared to other pairs, a serious problem arises in the inference with the standard posterior \eqref{eq:post} in the ordinal response model.
We will illustrate this through a simple example.

\begin{exm}
\label{exm:post}
We consider the Affairs data (\citealp{fair1978theory}).
The data contains 601 observations on three continuous, two binary, two ordered categorical, and two multinomial categorical variables.
The ordered categorical response variable is the self rating of marriage, and we standardized three continuous valued covariates, transform one ordered categorical covariate by the Likert sigma method, and create dummy variables from the multinomial categorical covariates.
Figure \ref{fg:grd_aff} shows the generalized residuals obtained by the Bayesian inference using the standard posterior \eqref{eq:post} with the probit link for the Affairs data.
The generalized residuals (\citealp{franses2001quantitative}) are calculated by
\begin{equation*}
\hat{e}_i = -\frac{g(\hat{\delta}_{y_i}-\bm{x}_i^\top\hat{\bm{\beta}}) - g(\hat{\delta}_{y_i-1}-\bm{x}_i^\top\hat{\bm{\beta}})}{G(\hat{\delta}_{y_i}-\bm{x}_i^\top\hat{\bm{\beta}}) - G(\hat{\delta}_{y_i-1}-\bm{x}_i^\top\hat{\bm{\beta}})},
\end{equation*}
where $\hat{\delta}_m$ ($m=1,2,3,4$) and $\hat{\beta}_j$ ($j=1,2,\ldots,18$) are the posterior means for the standard posterior.
The solid and dashed lines in the figure show the 95\% and 99\% intervals of the generalized residuals, respectively.
The large values of $\hat{e}_i$ may indicate the presence of outlying observations (\citealp{franses2001quantitative}).

Figure \ref{fg:post_aff} shows the standard posteriors for the coefficients of the ``affairs'' and ``age'' covariates using the original Affairs data and the modified data in which the absolute values of $\hat{e}_i$ larger than the 95\% interval are removed from the original data.
The posteriors using the original and modified data are different, i.e., the posterior inference with the standard posterior is affected by outliers and does not satisfy the posterior robustness (\citealp{desgagne2015robustness}).
We prove that this phenomenon occurs in the posterior inference using the standard posterior \eqref{eq:post} with arbitrary distribution function $G(\cdot)$ belonging to the parametric model (Theorem \ref{thm:stand_postrobust} in Section \ref{sec:stand_postrobust}).

\begin{figure}[H]
\begin{center}
\includegraphics[scale=0.45]{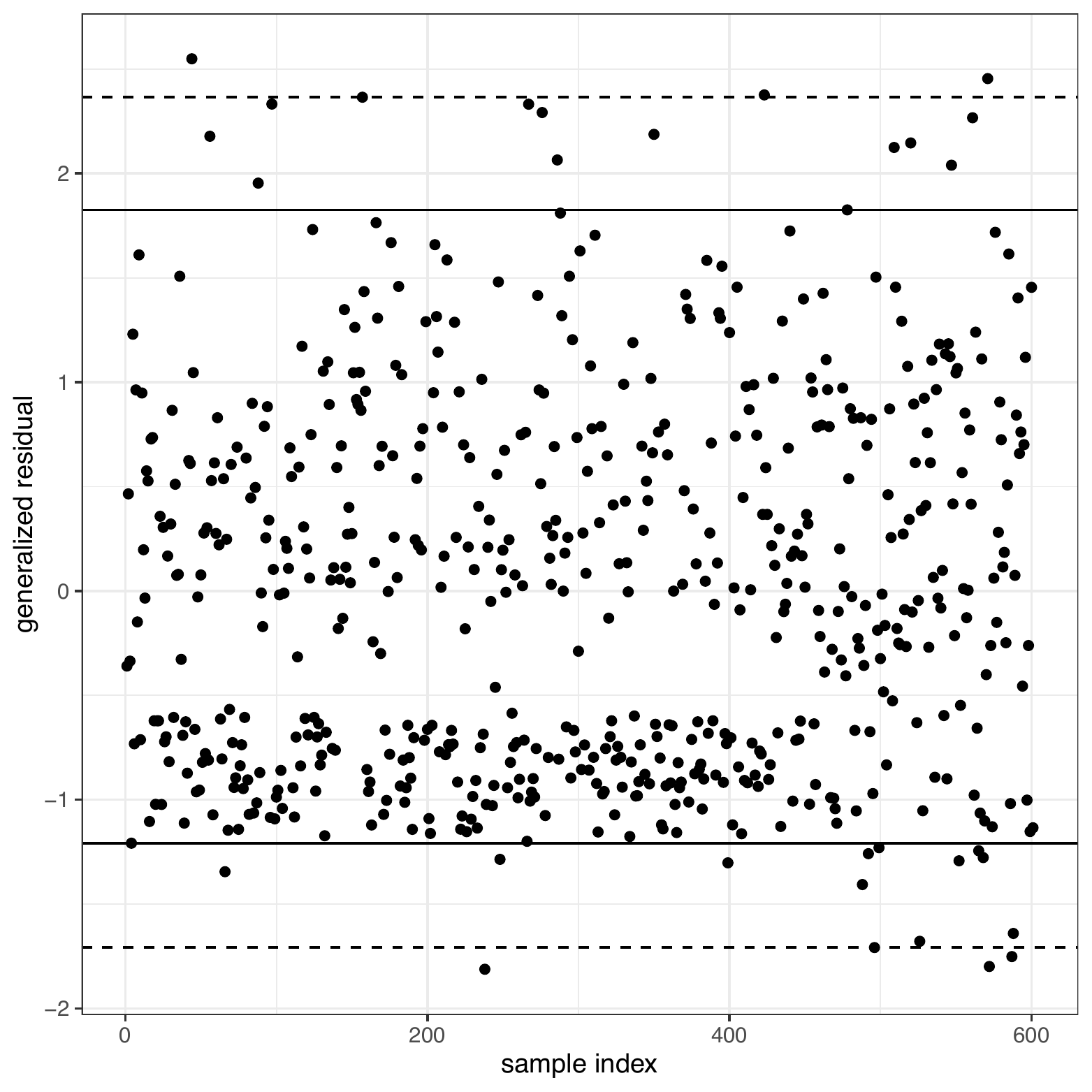}
\caption{The generalized residuals for the Affairs data. 
The solid and dashed lines show the 95\% and 99\% intervals of the generalized residuals, respectively.}
\label{fg:grd_aff}
\end{center}
\end{figure}

\begin{figure}[H]
\begin{center}
\includegraphics[scale=0.45]{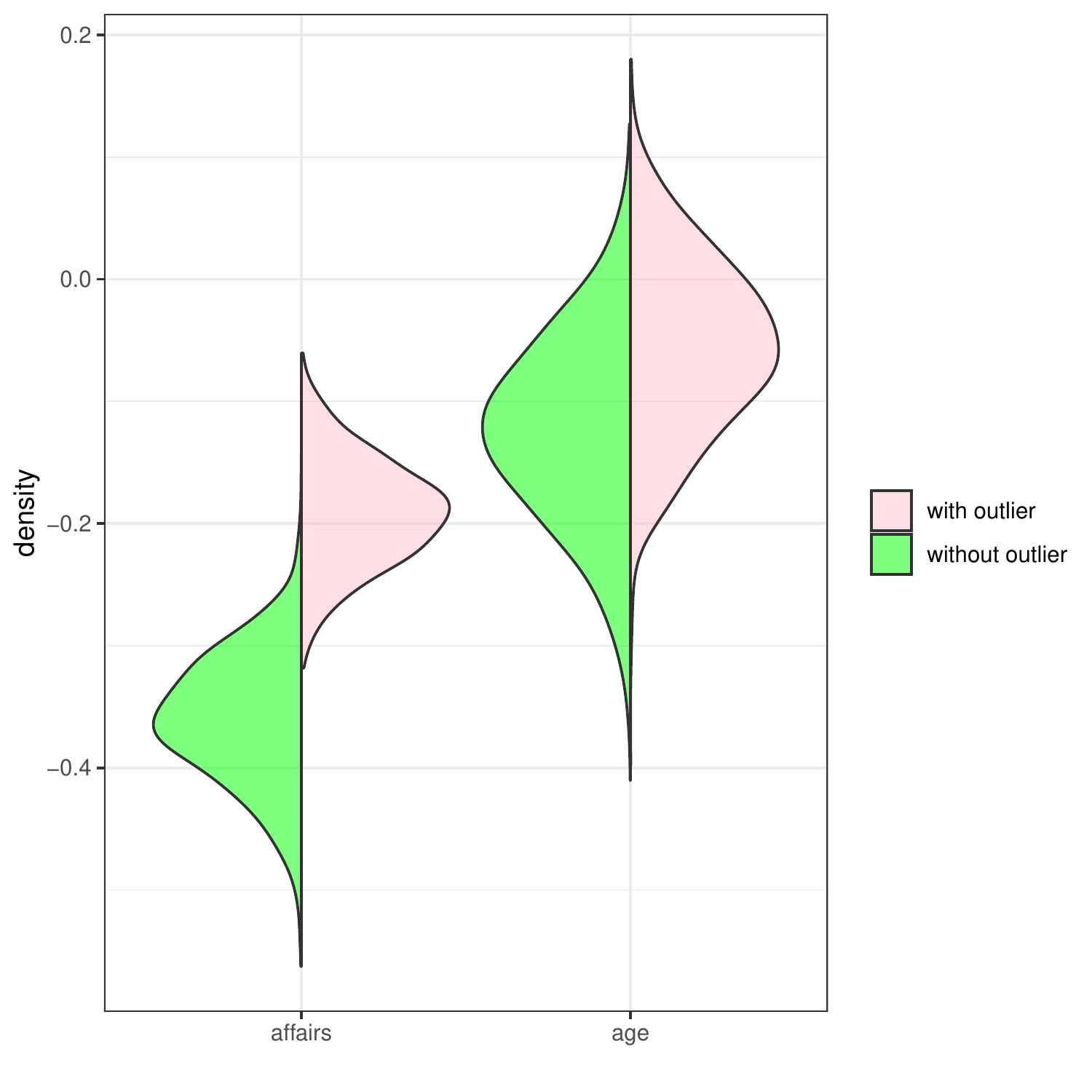}
\caption{The standard posteriors for the coefficients of the ``affairs'' (left) and ``age'' (right) covariates using the original Affairs data (green) and the modified data (pink) in which the absolute values of $\hat{e}_i$ larger than the 95\% interval are removed from the original data.}
\label{fg:post_aff}
\end{center}
\end{figure}

\end{exm}

%=sec:rborm=========================================================
\section{Robust bayesian ordinal response models via robust divergences}
\label{sec:rborm}
As described in Section \ref{sec:borm}, the Bayesian inference using the standard posterior \eqref{eq:post} is strongly affected by outliers.
To achieve robust Bayesian inference in the ordinal response model, we define general (synthetic) posteriors of the parameters based on the framework of the general (synthetic) posterior inference (\citealp{bissiri2016general, jewson2018robust, bhattacharya2019bayesian, miller2018robust, nakagawa2019robust, hashimoto2020robust}), by replacing the log-likelihood function with two robust divergences, the density-power (\citealp{basu1998robust}) and $\gamma$-divergences (\citealp{jones2001comparison, fujisawa2008robust}).
Section \ref{sec:density} defines a general posterior in the ordinal response model with the density-power divergence in the form of \cite{ghosh2016robust}.
Sections \ref{sec:gamma_HS} and \ref{sec:gamma_NH} define synthetic and general posteriors in the ordinal response model with \cite{hashimoto2020robust}'s and \cite{nakagawa2019robust}'s forms of the $\gamma$-divergence, respectively.
Section \ref{sec:post_comp} describes the method of the posterior computation based on our general posteriors.

\subsection{Case: density-power divergence} 
\label{sec:density}
Here, we use the following form of the density-power divergence in \cite{ghosh2016robust}.
\begin{equation*}
R_{DP}(D|\bm{\theta}) = \sum_{i=1}^n r_{DP}(y_i|\bm{x}_i;\bm{\theta}),
\end{equation*}
where
\begin{equation*}
r_{DP}(y_i|\bm{x}_i;\bm{\theta}) = \frac{1}{\alpha} f(y_i|\bm{x}_i;\bm{\theta})^\alpha - \frac{1}{1+\alpha} \sum_{y} f(y|\bm{x}_i;\bm{\theta})^{1+\alpha}
\end{equation*}
and $\alpha (>0)$ is a tuning parameter that controls the robustness.
Note that when $\alpha \to 0$, $R_{DP}(D|\bm{\theta})$ reduces to the log-likelihood function, and in the second term, the summation is used instead of the integral since $y$ is the discrete variable.

Then, using the density-power divergence $R_{DP}(D|\bm{\theta})$, we define the following general posterior in the ordinal response model, which will be referred to as the density-power posterior hereafter.
\begin{equation}
\label{eq:post_dp}
\begin{split}
p_{DP}(\bm{\theta}|D)
&= \frac{ p(\bm{\theta}) \exp\left\{ R_{DP}(D|\bm{\theta}) \right\} }{ p_{DP}(D) } \\
&= \frac{ p(\bm{\theta}) \prod_{i=1}^n \exp\left\{ r_{DP}(y_i|\bm{x}_i;\bm{\theta}) \right\} }{ \int p(\bm{\theta}) \prod_{i=1}^n \exp\left\{ r_{DP}(y_i|\bm{x}_i;\bm{\theta}) \right\} d\bm{\theta} },
\end{split}
\end{equation}
where $p_{DP}(D) = \int p(\bm{\theta}) \exp\left\{ R_{DP}(D|\bm{\theta}) \right\} d\bm{\theta}$.
Note that since the density-power posterior $p_{DP}(\bm{\theta}|D)$ is equivalent to the standard posterior \eqref{eq:post} when $\alpha\to0$, the density-power posterior may be regarded as a natural extension of the standard posterior \eqref{eq:post}.

We briefly confirm the proposed density-power posterior \eqref{eq:post_dp} is robust against outliers.
Suppose that the observation $(y_1,\bm{x}_1)$ is outlier compared to the other data, i.e., $|\bm{x}_1^\top\bm{\beta}|$ is large enough.
Then, since $G(\delta_{y_1}-\bm{x}_1^\top\bm{\beta}) - G(\delta_{y_1-1}-\bm{x}_1^\top\bm{\beta})$ takes a small value, the density-power divergence for the unit is expressed as
\begin{equation*}
r_{DP}(y_1|\bm{x}_1;\bm{\theta}) \approx -\frac{1}{1+\alpha}.
\end{equation*}
Thus, the density-power posterior \eqref{eq:post_dp} is expressed as
\begin{equation*}
p_{DP}(\bm{\theta}|D) \approx \frac{ p(\bm{\theta}) \prod_{i=2}^n \exp\left\{ r_{DP}(y_i|\bm{x}_i;\bm{\theta}) \right\} }{ \int p(\bm{\theta}) \prod_{i=2}^n \exp\left\{ r_{DP}(y_i|\bm{x}_i;\bm{\theta}) \right\} d\bm{\theta} }.
\end{equation*}
Namely, the density-power posterior \eqref{eq:post_dp} can be expressed by removing the outlier $(y_1,\bm{x}_1)$, and as a result, the influence of the outlier can be ignored in the posterior inference.

\subsection{Case: $\gamma$-divergence of the form in \cite{hashimoto2020robust}} 
\label{sec:gamma_HS}
Here, we use the following form of $\gamma$-divergence in \cite{hashimoto2020robust}.
\begin{equation*}
R_{\gamma(1)}(D|\bm{\theta}) = \frac{n}{\gamma} \log\left\{ \frac{\gamma}{n} \sum_{i=1}^n r_{\gamma}(y_i|\bm{x}_i;\bm{\theta}) \right\},
\end{equation*}
where
\begin{equation*}
r_{\gamma}(y_i|\bm{x}_i;\bm{\theta}) = \frac{1}{\gamma} \left( \frac{f(y_i|\bm{x}_i;\bm{\theta})}{||f(\cdot|\bm{x}_i;\bm{\theta})||_{\gamma+1}} \right)^{\gamma}, ~~ ||f(\cdot|\bm{x}_i;\bm{\theta})||_{\gamma+1} = \left( \sum_{y} f(y|\bm{x}_i;\bm{\theta})^{1+\gamma} \right)^{1/(1+\gamma)}
\end{equation*}
and $\gamma (>0)$ is a tuning parameter that controls the robustness.
Note that when $\gamma \to 0$, $R_{\gamma(1)}(D|\bm{\theta})$ reduces to the log-likelihood function, and in $||f(\cdot|\bm{x}_i;\bm{\theta})||_{\gamma+1}$, the summation is used instead of the integral since $y$ is the discrete variable.

Then, using the $\gamma$-divergence $R_{\gamma(1)}(D|\bm{\theta})$, we define the following synthetic posterior in the ordinal response model, which will be referred to as the $\gamma$-synthetic posterior hereafter.
\begin{equation}
\label{eq:post_g_HS}
\begin{split}
p_{\gamma(1)}(\bm{\theta}|D) &= \frac{ p(\bm{\theta}) \exp\left\{ R_{\gamma(1)}(D|\bm{\theta}) \right\} }{ p_{\gamma(1)}(D) } \\
&\propto p(\bm{\theta}) \left\{ \frac{\gamma}{n} \sum_{i=1}^n r_{\gamma}(y_i|\bm{x}_i;\bm{\theta}) \right\}^{n/\gamma}
\end{split}
\end{equation}
where $p_{\gamma(1)}(D) = \int p(\bm{\theta}) \exp\left\{ R_{\gamma(1)}(D|\bm{\theta}) \right\} d\bm{\theta}$.
Note that since the $\gamma$-synthetic posterior $p_{\gamma(1)}(\bm{\theta}|D)$ is equivalent to the standard posterior \eqref{eq:post} when $\gamma\to0$, the $\gamma$-synthetic posterior may be regarded as a natural extension of the standard posterior.

We briefly confirm the proposed $\gamma$-synthetic posterior \eqref{eq:post_g_HS} is robust against outliers.
Suppose that the observation $(y_1,\bm{x}_1)$ is outlier compared to the other data, i.e., $|\bm{x}_1^\top\bm{\beta}|$ is large enough.
Then, in terms of the similar manner of the previous subsection, since $G(\delta_{y_1}-\bm{x}_1^\top\bm{\beta}) - G(\delta_{y_1-1}-\bm{x}_1^\top\bm{\beta})$ takes a small value, the $\gamma$-divergence $R_{\gamma(1)}(D|\bm{\theta})$ for the unit is expressed as
\begin{equation*}
r_{\gamma}(y_1|\bm{x}_1;\bm{\theta}) \approx 0.
\end{equation*}
Thus, the $\gamma$-synthetic posterior \eqref{eq:post_g_HS} is expressed as
\begin{equation*}
p_{\gamma(1)}(\bm{\theta}|D) \approx p(\bm{\theta}) \left\{ \frac{\gamma}{n} \sum_{i=2}^n r_{\gamma}(y_i|\bm{x}_i;\bm{\theta}) \right\}^{n/\gamma}.
\end{equation*}
Namely, the $\gamma$-synthetic posterior \eqref{eq:post_g_HS} can be expressed by removing the outlier $(y_1,\bm{x}_1)$, and as a result, the influence of the outlier can be ignored in the posterior inference.

\subsection{Case: $\gamma$-divergence of the form in \cite{nakagawa2019robust}} 
\label{sec:gamma_NH}
Here, we use the following form of $\gamma$-divergence in \cite{nakagawa2019robust}.
\begin{align*}
R^*_{\gamma(2)}(D|\bm{\theta}) = \sum_{i=1}^n r_{\gamma}(y_i|\bm{x}_i;\bm{\theta}) - \frac{n}{\gamma} = R_{\gamma(2)}(D|\bm{\theta}) - \frac{n}{\gamma},
\end{align*}
where
\begin{equation*}
r_{\gamma}(y_i|\bm{x}_i;\bm{\theta}) = \frac{1}{\gamma} \left( \frac{f(y_i|\bm{x}_i;\bm{\theta})}{||f(\cdot|\bm{x}_i;\bm{\theta})||_{\gamma+1}} \right)^{\gamma}, ~~ ||f(\cdot|\bm{x}_i;\bm{\theta})||_{\gamma+1} = \left( \sum_{y} f(y|\bm{x}_i;\bm{\theta})^{1+\gamma} \right)^{1/(1+\gamma)}
\end{equation*}
and $\gamma (>0)$ is a tuning parameter that controls the robustness.
Note that when $\gamma \to 0$, $R^*_{\gamma(2)}(D|\bm{\theta})$ reduces to the log-likelihood function.

Then, using the $\gamma$-divergence $R^*_{\gamma(2)}(D|\bm{\theta})$, we define the following general posterior in the ordinal response model apart from \eqref{eq:post_g_HS}, which will be referred to as the $\gamma$-general posterior hereafter.
\begin{equation}
\label{eq:post_g_NH}
\begin{split}
p_{\gamma(2)}(\bm{\theta}|D) &= \frac{ p(\bm{\theta}) \exp\left\{ R^*_{\gamma(2)}(D|\bm{\theta}) \right\} }{ p^*_{\gamma(2)}(D) } \\
&= \frac{ p(\bm{\theta}) \exp\left\{ R_{\gamma(2)}(D|\bm{\theta}) \right\} }{ p_{\gamma(2)}(D) } = \frac{ p(\bm{\theta}) \prod_{i=1}^n \exp\left\{ r_{\gamma}(y_i|\bm{x}_i;\bm{\theta}) \right\} }{ \int p(\bm{\theta}) \prod_{i=1}^n \exp\left\{ r_{\gamma}(y_i|\bm{x}_i;\bm{\theta}) \right\} d\bm{\theta} },
\end{split}
\end{equation}
where $p^*_{\gamma(2)}(D) = \int p(\bm{\theta}) \exp\left\{ R^*_{\gamma(2)}(D|\bm{\theta}) \right\} d\bm{\theta}$ and $p_{\gamma(2)}(D) = \int p(\bm{\theta}) \exp\left\{ R_{\gamma(2)}(D|\bm{\theta}) \right\} d\bm{\theta}$.
Note that since the $\gamma$-general posterior $p_{\gamma(2)}(\bm{\theta}|D)$ is equivalent to the standard posterior \eqref{eq:post} when $\gamma\to0$, the $\gamma$-general posterior \eqref{eq:post_g_NH} may also be regarded as a natural extension of the standard posterior.

We briefly confirm the proposed $\gamma$-general posterior \eqref{eq:post_g_NH} is robust against outliers.
Suppose that the observation $(y_1,\bm{x}_1)$ is outlier compared to the other data, i.e., $|\bm{x}_1^\top\bm{\beta}|$ is large enough.
Then, in terms of the similar manner of the previous subsection, since $G(\delta_{y_1}-\bm{x}_1^\top\bm{\beta}) - G(\delta_{y_1-1}-\bm{x}_1^\top\bm{\beta})$ takes a small value, $r_{\gamma}(y_1|\bm{x}_1;\bm{\theta}) \approx 0$, and thus the $\gamma$-general posterior \eqref{eq:post_g_NH} is expressed as
\begin{equation*}
p_{\gamma(2)}(\bm{\theta}|D) \approx \frac{ p(\bm{\theta}) \prod_{i=2}^n \exp\left\{ r_{\gamma}(y_i|\bm{x}_i;\bm{\theta}) \right\} }{ \int p(\bm{\theta}) \prod_{i=2}^n \exp\left\{ r_{\gamma}(y_i|\bm{x}_i;\bm{\theta}) \right\} d\bm{\theta} }.
\end{equation*}
Namely, the $\gamma$-general posterior \eqref{eq:post_g_NH} can also be expressed by removing the outlier $(y_1,\bm{x}_1)$, and as a result, the infuluence of the outlier can be ignored in the posterior inference.

\begin{rem}
In the framework of the general posterior inference, the general Bayesian posterior is defined by
\begin{equation*}
p_{\rm GB}(\bm{\theta}|D) \propto p(\bm{\theta})\exp\left\{-w\ell(\bm{\theta},D)\right\},
\end{equation*}
where $w>0$ is the learning rate and $\ell(\bm{\theta},D)$ is the loss function defined additively for the observations $D$, i.e., $\ell(\bm{\theta},D) := \sum_{i=1}^n \ell(\bm{\theta},D_i)$.
Since the $\gamma$-synthetic posterior is not define by an additive loss function, we use the term ``synthetic''.
There have been many studies on how to choose the learning rate $w$, also known as the calibration weight or loss scale (\citealp{grunwald2017inconsistency, holmes2017assigning, lyddon2019general, syring2019calibrating, wu2023comparison}).
In this article, we set $w=1$ according to Section 2.3.2 of \cite{jewson2018robust}.
\end{rem}

\subsection{Posterior draws from proposed robust posteriors}
\label{sec:post_comp}
Consider how to draw the posterior samples from our proposed robust posteriors \eqref{eq:post_dp}, \eqref{eq:post_g_HS}, and \eqref{eq:post_g_NH}.
Since these posteriors do not have simple forms, the posterior sampling of $\bm{\theta}$ is intractable.
To solve this problem, we use the weighted likelihood bootstrap (WLB) method.
WLB is an approximate sampling method proposed by \cite{newton1994approximate}, whereby the posterior sampling from the posterior is obtained from the minimal solution of the weighted objective function.

The weighted objective functions in our proposed robust posteriors \eqref{eq:post_dp}, \eqref{eq:post_g_HS} and \eqref{eq:post_g_NH} are 
\begin{equation}
\label{eq:wl_g_NH}
WL_{DP}(\bm{\theta}) = - \sum_{i=1}^n s_i r_{DP}(y_i|\bm{x}_i;\bm{\theta}) - \log p(\bm{\theta}),
\end{equation}
\begin{equation}
\label{eq:wl_g_HS}
WL_{\gamma(1)}(\bm{\theta}) = - \frac{n}{\gamma} \log\left\{ \gamma \sum_{i=1}^n s_i r_{\gamma}(y_i|\bm{x}_i;\bm{\theta}) \right\} - \log p(\bm{\theta}),
\end{equation}
and
\begin{equation}
\label{eq:wl_g_NH}
WL_{\gamma(2)}(\bm{\theta}) = - \sum_{i=1}^n s_i r_{\gamma}(y_i|\bm{x}_i;\bm{\theta}) - \log p(\bm{\theta}),
\end{equation}
where $(s_1,s_2,\ldots,s_n) \sim {\rm Dirichlet}(1,1,\ldots,1)$, respectively.
Note that although the solution of the minimization problem in the weighted objective function is an approximate posterior sampling, \cite{lyddon2018nonparametric} and \cite{newton2021weighted} show that its approximation error is negligible when $n$ is sufficiently large.

This sampling method does not depend on the previous posterior samples of $\bm{\beta}$ and $\bm{\delta}$ to draw the current posterior sample of the parameters $\bm{\beta}$ and $\bm{\delta}$.
This means that the autocorrelation in each of $\bm{\beta}$ and $\bm{\delta}$ in this algorithm may be very small.
This is particularly useful in the sense that it solves the problem of autocorrelation among the posterior samples of $\bm{\delta}$, which is a bottleneck in \cite{albert1993bayesian}'s method.

\begin{rem}
Sampling from the posterior distributions \eqref{eq:post_dp}, \eqref{eq:post_g_HS}, and \eqref{eq:post_g_NH} can be implemented using the probabilistic programming language {\bf Stan} in addition to the method using the WLB.
However, it uses the Hamiltonian Monte Carlo method, which is one of the MCMC methods, and depending on the structure of the data, the computational performance may be poor in terms of the number of effective samplings.
In particular, the performance tends to deteriorate as the number of dimensions $p$ increases.
We will illustrate this through a simple example.
Suppose that an ordered categorical data $y_i$ is generated from \eqref{eq:lvm} with $n=200$ and $p=20$, where $\bm{x}_i$ is generated from $N_p(\bm{0}, \bm{I}_p)$ with the $p\times p$ identity matrix $\bm{I}_p$, $\bm{\beta}$ is generated from ${\rm Unif}(-2,2)$, $\ep_i \overset{i.i.d.}{\sim} N(0,1)$, and $\bm{\delta} = (-2.3, -0.7, 0.7, 2.3)$.
Note that this is the ordinal probit model, that is, the link function in the ordinal response model is the probit link.
We obtain the 2000 posterior draws with the WLB.
In the method with the Stan, 500 of the 2500 posterior draws are discarded as burn-in and the rest are used.

Figures \ref{fg:wlb_stan_beta1} and \ref{fg:wlb_stan_delta1} show trace plots and autocorrelations for the draws of $\beta_1$ and $\delta_1$ from the density-power and $\gamma$-general posteriors \eqref{eq:post_dp}, \eqref{eq:post_g_NH} with the WLB and Stan.
As can be seen from these figures, in the case of the WLB, solving the minimization problem of weighted objective function, independent posterior draws can be obtained, hence the mixing is quite well and the autocorrelation is negligible.
Contrarily, the Stan algorithm causes poor mixing and quite high autocorrelation.
More unfortunately, as p grows, the posterior draws with the Stan are sampled far away from the true values of the parameters ($\beta_1=-0.94$, $\delta_1=-2.3$ in this simulation).
Even increasing the length of MCMC sequences does not resolve this problem at all, and in our experience, such a situation is more likely to occur when $p$ is larger than 15.
In this simulation, we used 0.5 as the value of the tuning parameter, but other values did not change the results of the mixing and autocorrelations.
The results for the $\gamma$-synthetic posterior \eqref{eq:post_g_HS} are not different from those for the $\gamma$-general posterior \eqref{eq:post_g_NH}, so we omit the details.

\begin{figure}[H]
  \begin{minipage}[b]{0.5\linewidth}
    \centering
    \includegraphics[keepaspectratio, width=\columnwidth]{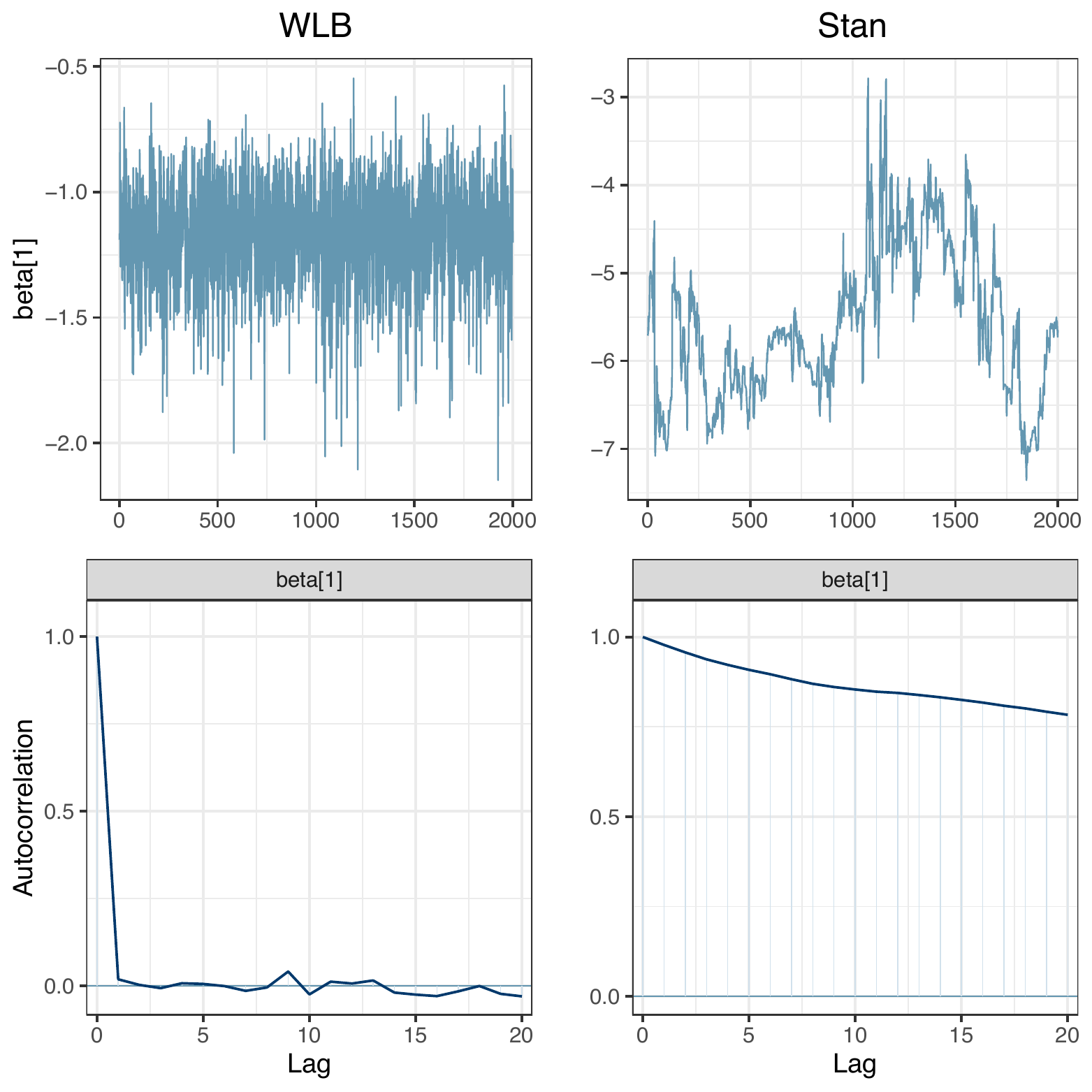}
    \subcaption{Density-power posterior \eqref{eq:post_dp}}
  \end{minipage}
  \begin{minipage}[b]{0.5\linewidth}
    \centering
    \includegraphics[keepaspectratio, width=\columnwidth]{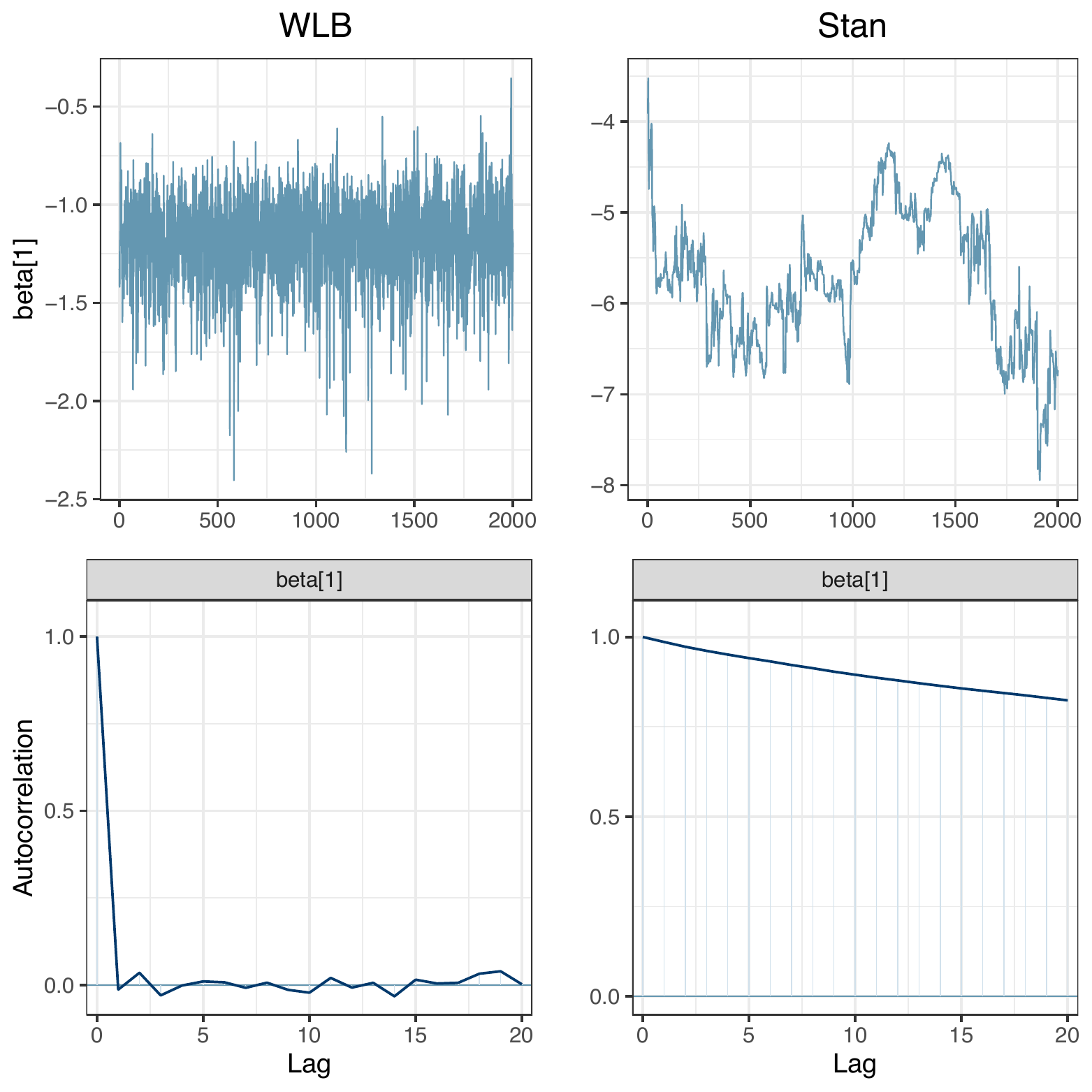}
    \subcaption{$\gamma$-general posterior \eqref{eq:post_g_NH}}
  \end{minipage}
  \caption{Trace plots and autocorrelations for the draws of $\beta_1$ from the density-power and $\gamma$-general posteriors with the WLB and Stan.}
  \label{fg:wlb_stan_beta1}
\end{figure} %

\begin{figure}[H]
  \begin{minipage}[b]{0.5\linewidth}
    \centering
    \includegraphics[keepaspectratio, width=\columnwidth]{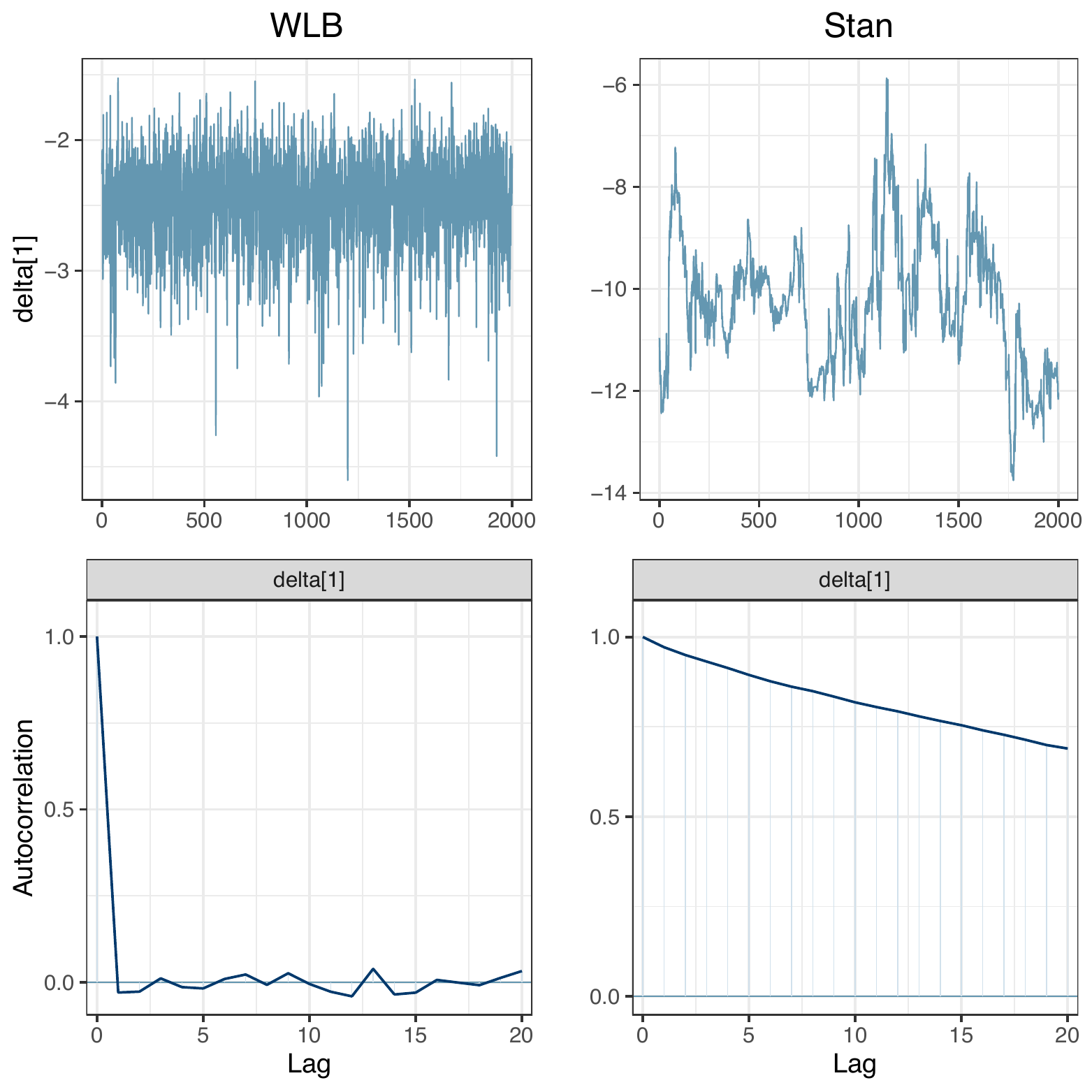}
    \subcaption{Density-power posterior \eqref{eq:post_dp}}
  \end{minipage}
  \begin{minipage}[b]{0.5\linewidth}
    \centering
    \includegraphics[keepaspectratio, width=\columnwidth]{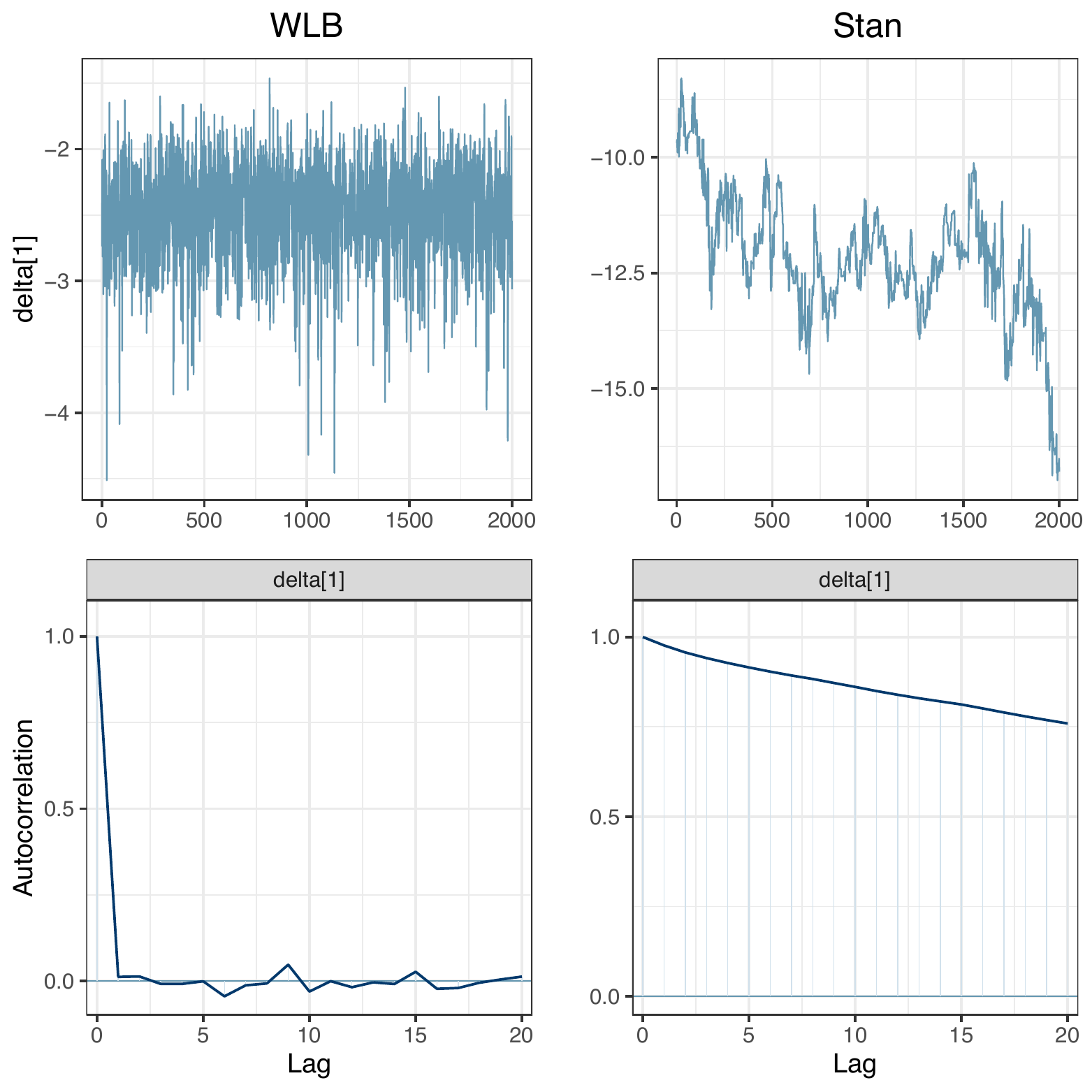}
    \subcaption{$\gamma$-general posterior \eqref{eq:post_g_NH}}
  \end{minipage}
  \caption{Trace plots and autocorrelations for the draws of $\delta_1$ from the density-power and $\gamma$-general posteriors with the WLB and Stan.}
  \label{fg:wlb_stan_delta1}
\end{figure} %
\end{rem}

\section{Posterior robustness of standard and proposed posteriors}
\label{sec:post_robust}  
The previous section defined the general (synthetic) posteriors \eqref{eq:post_dp}, \eqref{eq:post_g_HS}, \eqref{eq:post_g_NH} in the ordinal response model using the density-power and $\gamma$-divergences, and developed the robust Bayesian ordinal response model.
In this section, we prove that our three robust posteriors \eqref{eq:post_dp}, \eqref{eq:post_g_HS}, \eqref{eq:post_g_NH} are robust against outliers in terms of the posterior robustness (\citealp{desgagne2015robustness}), including the standard posterior \eqref{eq:post}.
Before proving the posterior robustness, we redefine the notation for outliers with the similar manner of \cite{desgagne2019bayesian}.

The set of indices for $n$ observations, $\{1,2,\ldots,n\}$, is divided into two mutually disjoint subsets, $\mathcal{K}$ and $\mathcal{L}$, and let $\mathcal{K}$ denote the set of indices that are not outliers and $\mathcal{L}$ denote the set of outlier indices.
Note that $\mathcal{K} \cup \mathcal{L} = \{1,2,\ldots,n\}$ and $\mathcal{K} \cap \mathcal{L}$ is the empty set.
Let $D_i = (y_i, \bm{x}_i)$ denote the $i$-th observation and $D=\{D_1,D_2,\ldots,D_n\}$ be the set of the observed data.
The set of the non-outlying observations is defined by $D^* = \{D_i | i \in \mathcal{K}\}$.
Then the (non-)outliers for the observed values are defined as
\begin{equation*}
u_i := \delta_{y_i} - \bm{x}_i^\top\bm{\beta} = 
\begin{cases}
a_i, & (i \in \mathcal{K}), \\
a_i + b_i \omega, & (i \in \mathcal{L}),
\end{cases}
\end{equation*}
where $a_i \in \mathbb{R}$, $b_i \neq 0$ which describes the direction of outlier, and $\omega > 0$.
If $\omega$ is sufficiently large, the value of $u_i$ for $i\in\mathcal{L}$ is extremely large, either positively or negatively due to $b_i$, i.e., the observation $D_i$ is outlier.
Note that a large value of $u_i$ means a large absolute value of generalized residuals (see Example \ref{exm:post}).

\subsection{Case: standard posterior}
\label{sec:stand_postrobust}
For the standard posterior \eqref{eq:post} in the ordinal response model, the following theorem on the posterior robustness holds.
\begin{thm}
\label{thm:stand_postrobust}
Assume that the probability density function $g(\cdot)$ of the distribution $G(\cdot)$ is proper.
Then, 
\begin{equation*}
\lim_{\omega\to\infty} p(\bm{\theta}|D) \neq p(\bm{\theta}|D^*)
\end{equation*}
holds.
\end{thm}
\noindent
The proof of Theorem \ref{thm:stand_postrobust}, see Appendix \ref{app:stand_postrobust}.

Theorem \ref{thm:stand_postrobust} implies that the standard posterior in the ordinal response model cannot satisfy the posterior robustness, i.e., the posterior inference without removing the influence of outliers is not possible with the conventional Bayesian ordinal response model.
In the inferences of the location and scale parameters, and the linear regression, it is several well-known results that the posterior robustness can be satisfied by using the heavy-tailed distributions (\citealp{o2012bayesian}).
For example, \cite{gagnon2020new} and \cite{hamura2022log} prove that the standard posterior in the linear regression satisfies the posterior robustness by using the extremely heavy tail distribution whose order of the tail is same as $(x\log x)^{-1}$ (e.g., the log-Pareto tailed normal and extremely heavily-tailed distributions).
Interestingly, however, the standard posterior in the ordinal response model requires the distribution of $\ep_i$ to have the tail of order smaller than $1/x$, so the standard posterior \eqref{eq:post} cannot satisfy the posterior robustness even if the extremely heavy tail distribution is used, which can achieve the posterior robustness in the linear regression.
It is also important to note that the standard posterior \eqref{eq:post} with the distribution that satisfies the regularly varying (\citealp{o2012bayesian}), such as the Student's $t$-distribution and the Cauchy distribution, cannot satisfy the posterior robustness.

\subsection{Case: general (synthetic) posteriors}
\label{sec:general_postrobust}
For the proposed general (synthetic) posteriors \eqref{eq:post_dp}, \eqref{eq:post_g_HS}, \eqref{eq:post_g_NH} in the ordinal response model, the following theorems on the posterior robustness hold.
\begin{thm}
\label{thm:dp_postrobust}
Assume that $p(\bm{\theta})$ is proper. 
Then the density-power posterior \eqref{eq:post_dp} satisfies the posterior robustness, i.e.,
\begin{equation*}
\lim_{\omega\to\infty} p_{DP}(\bm{\theta}|D) = p_{DP}(\bm{\theta}|D^*),
\end{equation*}
where $p_{DP}(\bm{\theta}|D)$ are the density-power posterior that is conditioned by the full dataset $D$, and $p_{DP}(\bm{\theta}|D^*)$ are the density-power posterior that is conditioned by the dataset without the outliers $D^*$.
\end{thm}

\begin{thm}
\label{thm:gamma_HS_postrobust}
Assume that $p(\bm{\theta})$ is proper. 
Then the $\gamma$-synthetic posterior \eqref{eq:post_g_HS} satisfies the posterior robustness, i.e.,
\begin{equation*}
\lim_{\omega\to\infty} p_{\gamma(1)}(\bm{\theta}|D) = p_{\gamma(1)}(\bm{\theta}|D^*),
\end{equation*}
where $p_{\gamma(1)}(\bm{\theta}|D)$ are the synthetic $\gamma$-posterior that is conditioned by the full dataset $D$, and $p_{\gamma(1)}(\bm{\theta}|D^*)$ are the synthetic $\gamma$-posterior that is conditioned by the dataset without the outliers $D^*$.
\end{thm}

\begin{thm}
\label{thm:gamma_NH_postrobust}
Assume that $p(\bm{\theta})$ is proper. 
Then the $\gamma$-general posterior \eqref{eq:post_g_NH} satisfies the posterior robustness, i.e.,
\begin{equation*}
\lim_{\omega\to\infty} p_{\gamma(2)}(\bm{\theta}|D) = p_{\gamma(2)}(\bm{\theta}|D^*),
\end{equation*}
where $p_{\gamma(2)}(\bm{\theta}|D)$ are the general $\gamma$-posterior that is conditioned by the full dataset $D$, and $p_{\gamma(2)}(\bm{\theta}|D^*)$ are the general $\gamma$-posterior that is conditioned by the dataset without the outliers $D^*$.
\end{thm}
\noindent
The proofs of Theorems \ref{thm:dp_postrobust}, \ref{thm:gamma_HS_postrobust}, and \ref{thm:gamma_NH_postrobust}, see Appendix \ref{app:dp_postrobust}, \ref{app:gamma_HS_postrobust}, and \ref{app:gamma_NH_postrobust}, respectively.

While the standard posterior \eqref{eq:post} in the conventional Bayesian ordinal response model could not satisfy the posterior robustness, the general (synthetic) posteriors \eqref{eq:post_dp}, \eqref{eq:post_g_HS}, \eqref{eq:post_g_NH} in our state-of-the-art robust Bayesian ordinal response model can satisfy the posterior robustness for any distribution function $G(\cdot)$.
Namely, the Bayesian inference with these general (synthetic) posteriors gives a choice of the link function to those analyzing data that may (or may not) contain outliers, and provides robust and flexible results against outliers.

\section{Numerical experiments}
\label{sec:ne}

\subsection{Bayesian robustness property}
\label{sec:ne_robust}
This section demonstrates that our proposed robust Bayesian ordinal response model is robust against outliers through a numerical experiment using the robustness index proposed by \cite{kurtek2015bayesian}.
The robustness index represents the influence of each individual data on the inference, i.e., the value of the index should be small for heterogeneous individual data.

In this numerical experiment, we employ the following simple ordinal response model as the data generation process.
\begin{equation*}
y = m \Leftrightarrow \delta_{m-1} < z \leq \delta_{m}, ~~ z = \beta x + \ep,
\end{equation*}
where $\beta=0.7$, $\bm{\delta} = (-1.6, 0, 1.6)$, $x$ is a value that ticks from $-5$ to 5 in increments of 0.05, and $\ep \sim N(0,1)$.
As comparison methods with our proposed robust Bayesian ordinal response model, we use the conventional Bayesian ordinal response model with the probit and cauchit (the distribution function of the Cauchy distribution) links as the link function.
In our proposed robust Bayesian ordinal response model, the probit link is used as the link function, and both the density-power and $\gamma$-divergences set the tuning parameter of the robust divergences to 0.3, 0.5, 0.7, and 1.0.
To run the conventional Bayesian ordinal response model, we use the function \texttt{stan\_polr} of the {\bf rstanarm} package in the {\bf R} programming language.
For all Bayesian inferences performed in this numerical experiment, 2500 samples are taken, of which 500 are burn-in.

\begin{figure}[H]
\begin{center}
\includegraphics[width=\columnwidth]{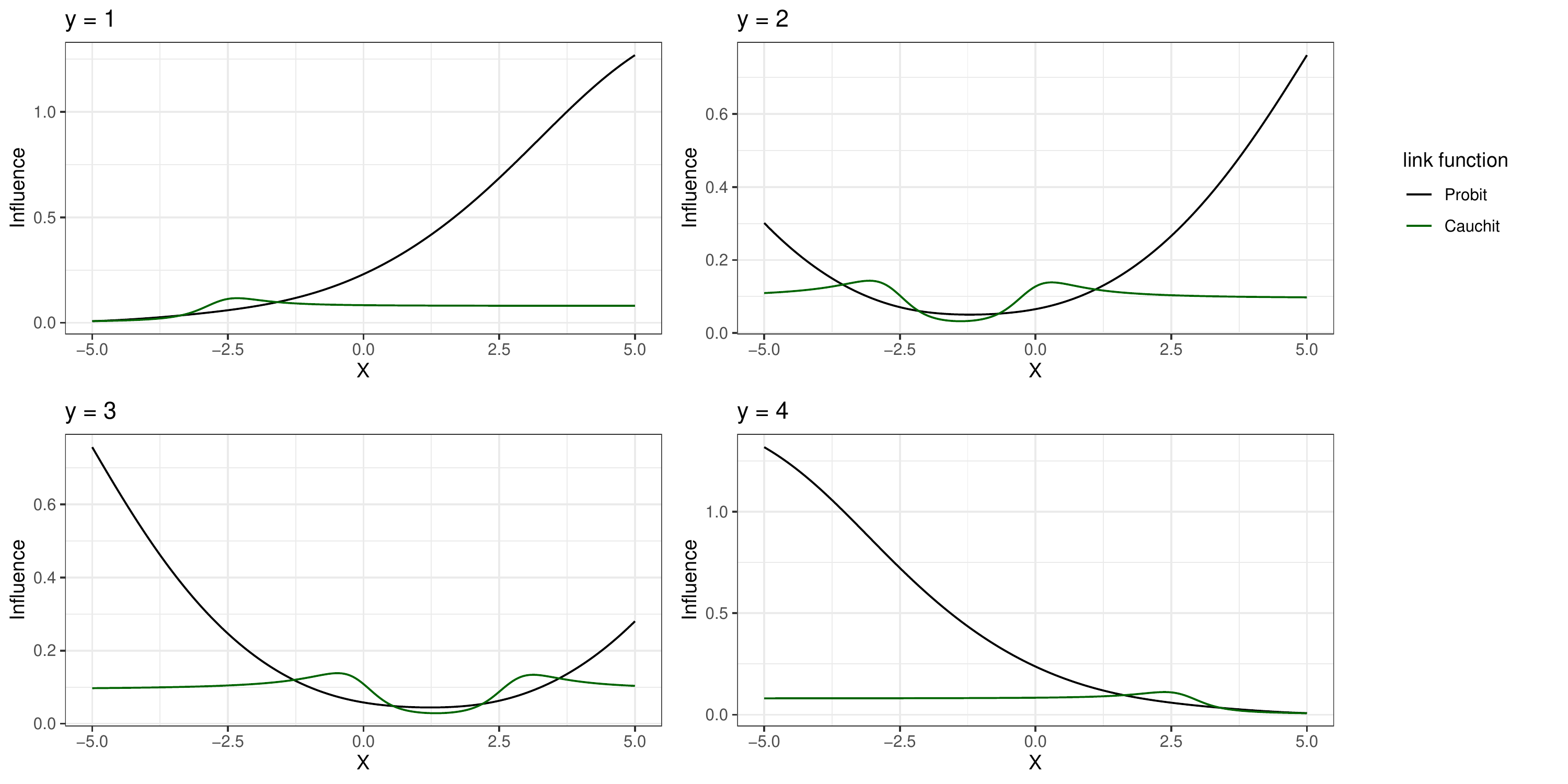}
\caption{Plots of the values of the robustness index (\citealp{kurtek2015bayesian}) when using the conventional Bayesian ordinal response model with the probit (black) and cauchit (green) links.}
\label{fg:if_KL}
\end{center}
\end{figure}

Figure \ref{fg:if_KL} shows the values of robustness index of \cite{kurtek2015bayesian} when using the conventional Bayesian ordinal response model with the probit (black line) and cauchit (green line) links.
It can be seen that the values of the index is considerably large for heterogeneous individual data in the conventional Bayesian ordinal response model with the probit link.
Here, the heterogeneous data means, for example, $y=1$ (upper left panel of Figure \ref{fg:if_KL}), where $x$ may be negative in the latent variable model used in this numerical experiment, so that the data where $y=1$ and $x$ is positive may be said to be heterogeneous.
In contrast to the case with the probit link, the values of the robustness index in the conventional Bayesian ordinal response model with the cauchit link are small even for heterogeneous individual data, but the value is not zero.
This means that the conventional Bayesian ordinal response model with the cauchit link can suppress the influence of heterogeneous data to some extent, but cannot eliminate it completely.

\begin{figure}[H]
\begin{center}
\includegraphics[width=\columnwidth]{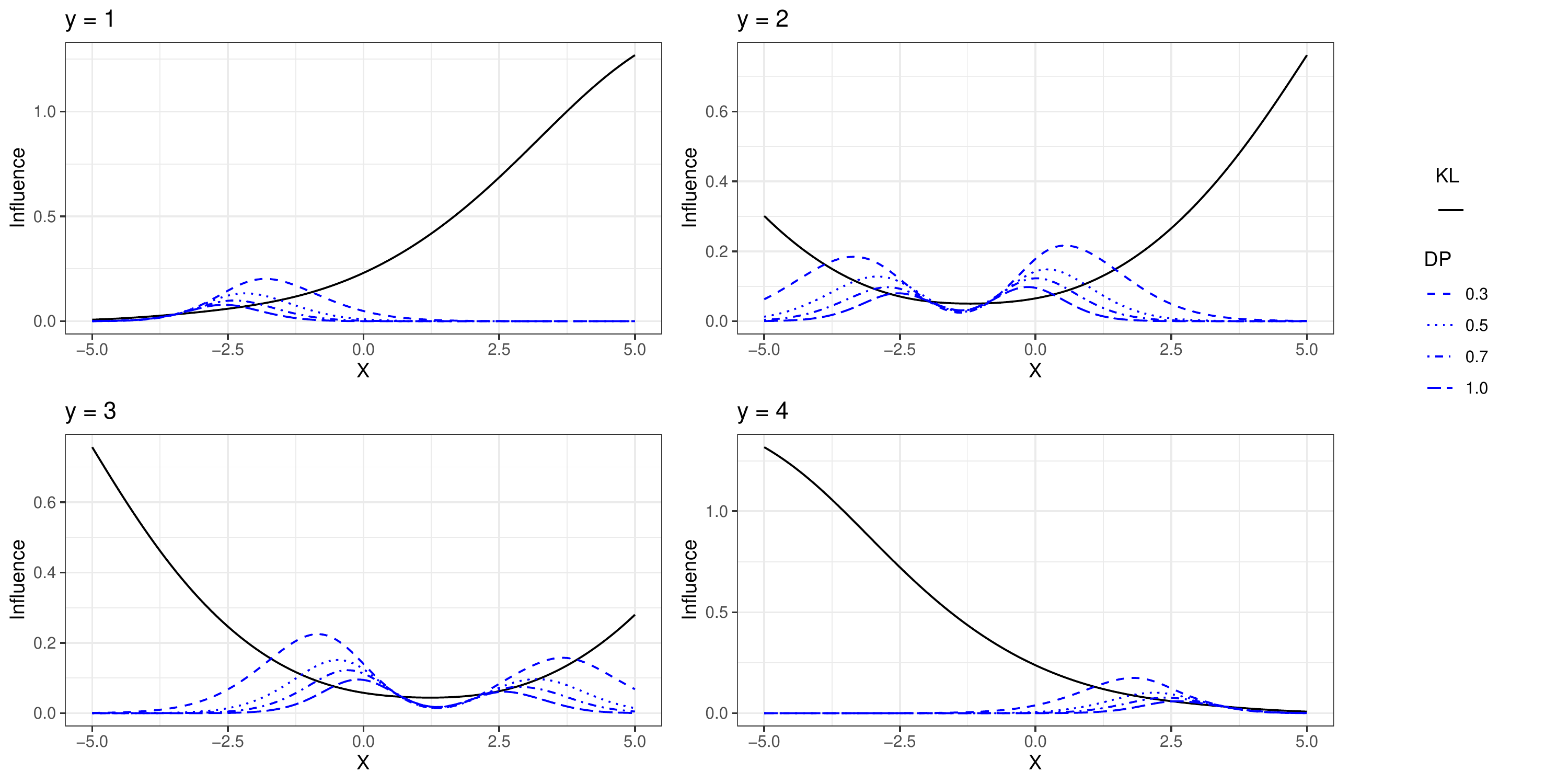}
\caption{Plots of the values of the robustness index using the proposed robust Bayesian ordinal response model with the density-power divergence (blue).
Each blue line corresponds to the values of the robustness index for different values of the tuning parameter of the density-power divergence.
The black line represents the values of the index when using the conventional Bayesian ordinal response model with the probit link (posterior with the Kullback-Leibler loss).}
\label{fg:if_DP}
\end{center}
\end{figure}

\begin{figure}[H]
\begin{center}
\includegraphics[width=\columnwidth]{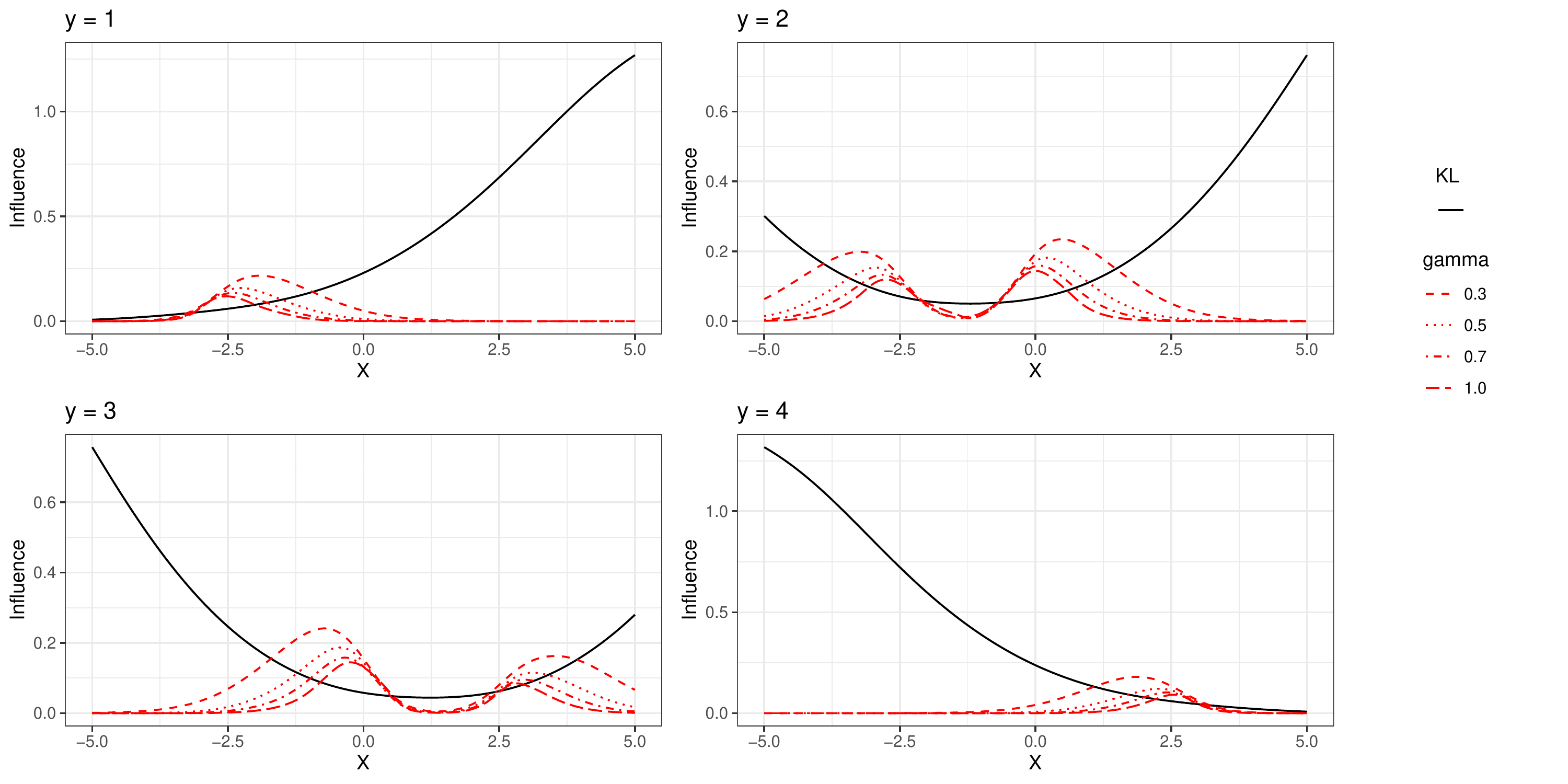}
\caption{Plots of the values of the robustness index using the proposed robust Bayesian ordinal response model with the $\gamma$-divergence (red).
Each red line corresponds to the values of the robustness index for different values of the tuning parameter of the $\gamma$-divergence.
The black line represents the values of the index when using the conventional Bayesian ordinal response model with the probit link (posterior with the Kullback-Leibler loss).}
\label{fg:if_g_NH}
\end{center}
\end{figure}

Figures \ref{fg:if_DP} and \ref{fg:if_g_NH} show the values of the robustness index when using the density-power posterior \eqref{eq:post_dp} and $\gamma$-general posterior \eqref{eq:post_g_NH} in our proposed robust Bayesian ordinal response model, respectively. 
The black line represents the values of the index when using the conventional Bayesian ordinal response model with the probit link (posterior with the Kullback-Leibler loss).
The results using the $\gamma$-synthetic posterior \eqref{eq:post_g_HS} show almost the same behavior as those using the $\gamma$-general posterior \eqref{eq:post_g_NH}, so the output is omitted in this numerical experiment.
As can be seen from these results, the values of the index in our proposed robust Bayesian ordinal response model with the robust divergences approaches zero for heterogeneous data.
Namely, our method not only suppresses the influence of heterogeneous data, but also completely eliminates it, indicating that our method can achieve a more robust inference than the conventional Bayesian ordinal response model.

\subsection{Simulation studies}
\label{sec:ne_simu}
This section evaluates the performance of our proposed robust Bayesian ordinal response model through this numerical experiments.
In order to compare with the conventional Bayesian ordinal response model, we perform some simulation studies referring to Section 5 of \cite{scalera2021robust}.
We employ the following ordinal response model as the data generation process.
\begin{equation*}
y = m \Leftrightarrow \delta_{m-1} < z \leq \delta_{m}, ~~ z = x\beta_1 + d\beta_2 + xd\beta_3 + \ep,
\end{equation*}
where $x\sim (1-\rho)N(0,1)+\rho N(20,1)$ with outlier ratio $\rho=(0,0.05,0.10,0.15,0.20)$, $d\sim {\rm Bernoulli}(0.25)$, and $x$ and $d$ are mutually independent.
The random variable $\ep$ is distributed as the standard normal, logistic, and Gumbel, and in this numerical experiment, we use the link function corresponding to the error distribution, for example, if $\ep$ has the standard normal distribution, then we use the probit link.
This is because it is well known that the misspecification of the link function can cause a substantial bias in the inference.
The development of a doubly robust Bayesian ordinal response model for both outliers and the link misspecification is a future work.
We set the values of the regression coefficients $\bm{\beta}$ and cutoffs $\bm{\delta}$ as follows according to the error distribution.
\begin{itemize}
\item When $\ep$ is distributed as the standard normal:
	\begin{equation*}
	(\beta_1, \beta_2, \beta_3) = (2.5, 1.2, 0.7) ~~ \mbox{and} ~~ (\delta_1, \delta_2, \delta_3, \delta_4) = (-3.0, -0.7, 1.6, 3.9).
	\end{equation*}
\item When $\ep$ is distributed as the standard logistic:
	\begin{equation*}
	(\beta_1, \beta_2, \beta_3) = (2.5, 1.2, 0.7) ~~ \mbox{and} ~~ (\delta_1, \delta_2, \delta_3, \delta_4) = (-3.3, -0.8, 1.7, 4.2).
	\end{equation*}
\item When $\ep$ is distributed as the Gumbel:
	\begin{equation*}
	(\beta_1, \beta_2, \beta_3) = (2.5, 1.2, 0.7) ~~ \mbox{and} ~~ (\delta_1, \delta_2, \delta_3, \delta_4) = (-2.9, 1.0, 2.9, 4.8).
	\end{equation*}
\end{itemize}
With the above settings, we apply our proposed Bayesian ordinal robust model with the density-power posterior (RBOR DPP), $\gamma$-synthetic posterior (GSP), and $\gamma$-general posterior (GGP) to these data and generate 2000 posterior samples using the WLB method.
The tuning parameters ($\alpha$ and $\gamma$) for the density-power and $\gamma$-divergences are set to 0.3 and 0.5.
For comparison, we apply the non-robust Bayesian ordinal response model with the standard posterior (BOR) and generate 2500 posterior samples using the function \texttt{stan\_polr} of the {\bf rstanarm} package in the {\bf R}, 500 of which are burn-in.
The Bayesian inference with the standard posterior distribution using the cauchit link (c-BOR), which is the distribution function of the Cauchy distribution, is also included in the comparison with our methods.
This is because indeed, as mentioned earlier, the cauchit link does not satisfy the posterior robustness, and the misspecification of the link function can cause a substantial bias in the parameter estimation, but from the numerical experiment using the robustness index in Section \ref{sec:ne_robust}, the Bayesian inference using the cauchit link seems to give more robust inference results against outliers than using the probit link.
For the point estimates of $\theta_j$, we use their posterior means $\hat{\theta}_j$ and evaluate their performance in terms of mean squared error (MSE) defined by $q^{-1} \sum_{j=1}^q (\hat{\theta}_j-\theta_j)^2$ with the number of parameters.
We also use coverage probabilities (CP) defined by $q^{-1} \sum_{j=1}^q I(\theta_j \in {\rm CI}_j)$ to evaluate $95\%$ credible intervals (CI) of $\theta_k$.
The 200 observations are generated by the above data generation processes and these values are computed for each of the 100 replicates of the artificial dataset and averaged.

\begin{figure}[H]
  \begin{minipage}[b]{0.49\linewidth}
    \centering
    \includegraphics[keepaspectratio, width=\columnwidth]{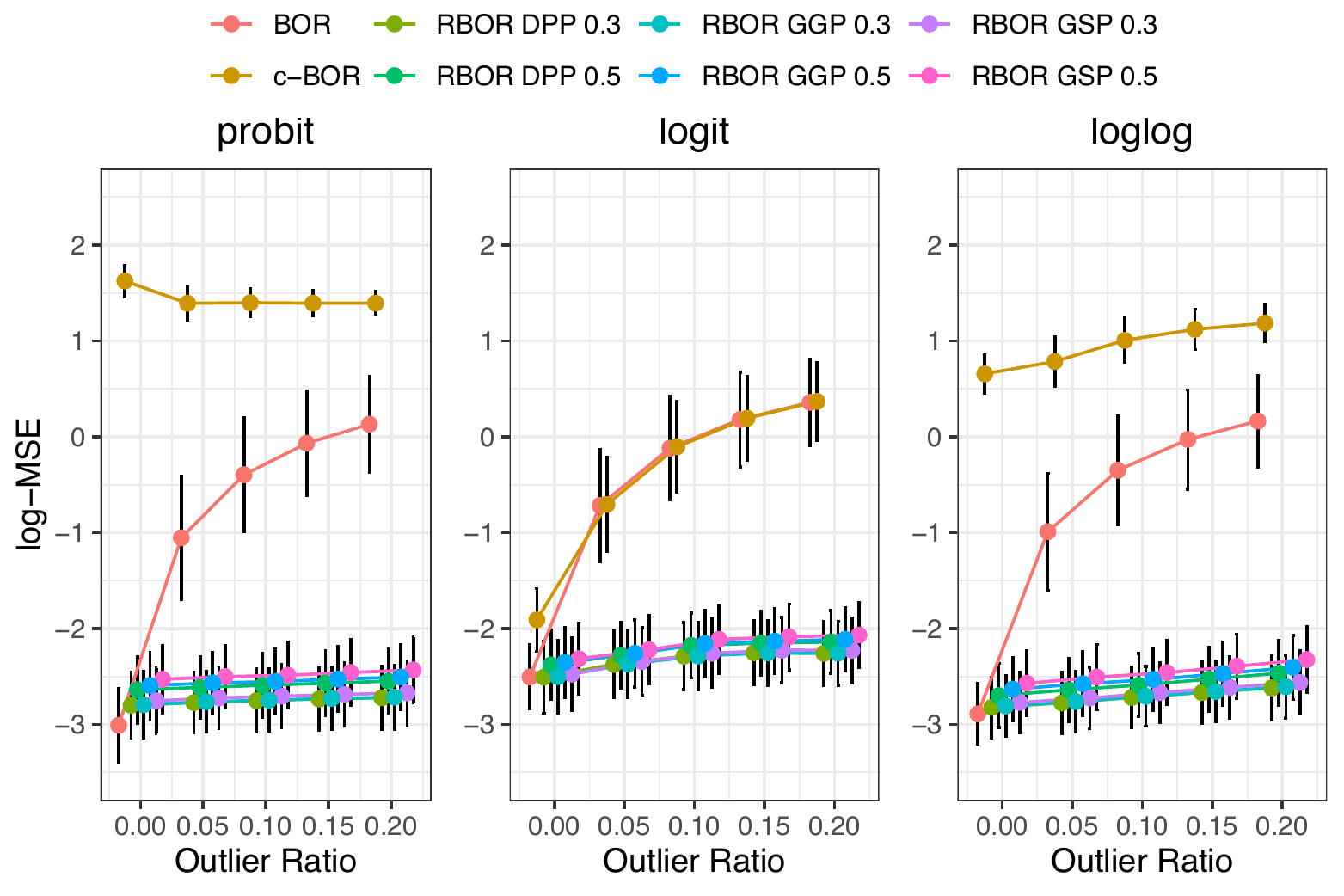}
    \subcaption{Coefficient parameters $\bm{\beta}$}
  \end{minipage}
  \begin{minipage}[b]{0.49\linewidth}
    \centering
    \includegraphics[keepaspectratio, width=\columnwidth]{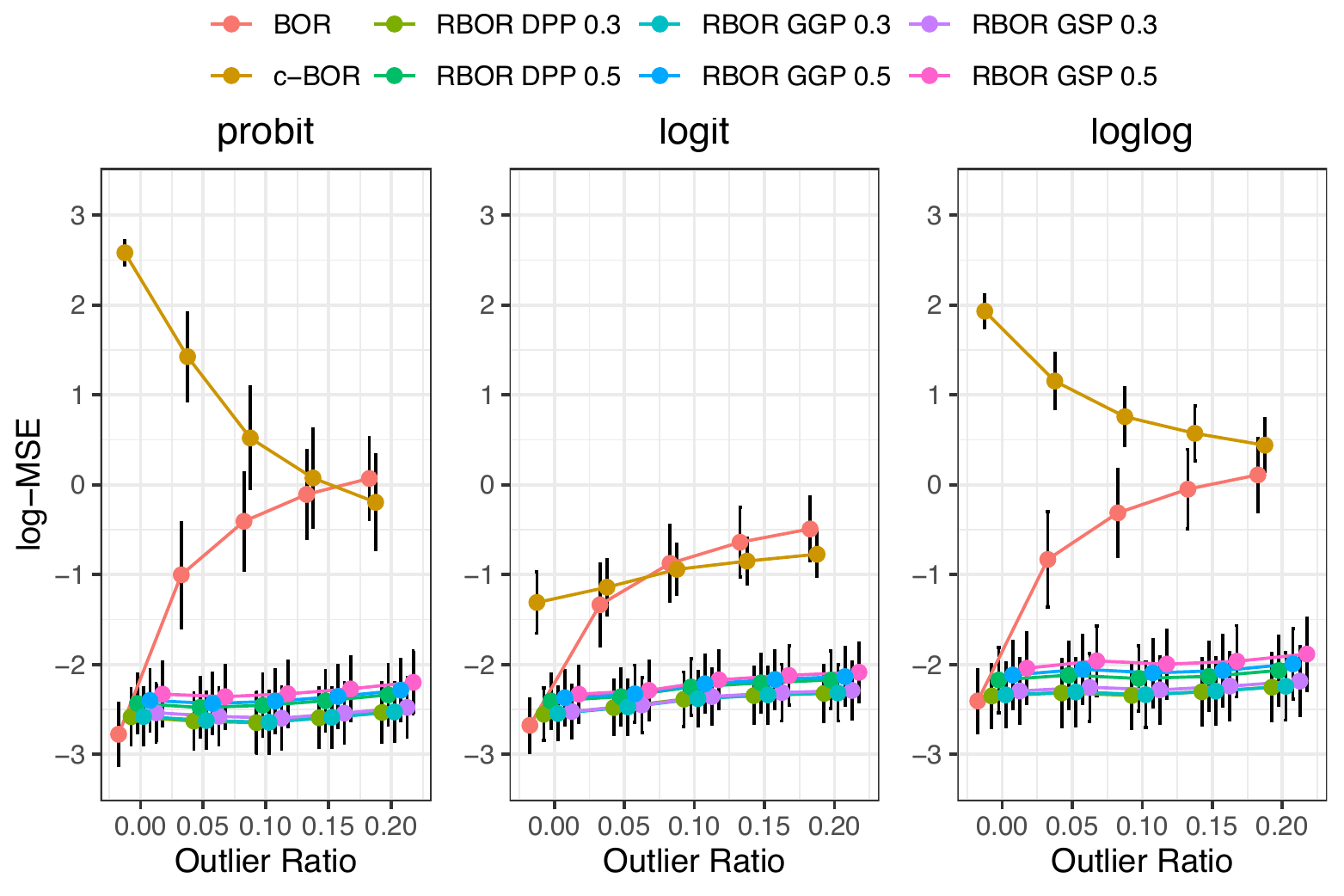}
    \subcaption{Cutoff parameters $\bm{\delta}$}
  \end{minipage}
  \caption{The log-MSE values with estimated Monte Carlo error bars of each method for each outlier ratio}
  \label{fg:ne_or}
\end{figure}

Figure \ref{fg:ne_or} plots the logarithm of MSE (log-MSE) values with estimated Monte Carlo error bars of each method for each outlier ratio.
In the absence of outliers, our proposed robust methods (RBOR DPP, GSP, and GGP) perform as well as the conventional BOR methods.
When the outlier ratio increases, the performance of BOR deteriorates, while the performance of our proposed robust methods remains almost the same.
The c-BOR method using the cauchit link is found to have bounded values of the robustness index of \cite{kurtek2015bayesian} as shown in Figure \ref{fg:if_KL} of Section \ref{sec:ne_robust}, but the parameter estimation is adversely affected due to the misspecification of the link function.
Tables \ref{tb:cp_beta} and \ref{tb:cp_delta} present the results for the CP of the 95\% credible intervals regarding the interval estimation for the coefficient parameters $\bm{\beta}$ and the cutoff parameters $\bm{\delta}$.
These results show that our proposed robust methods perform as well as the conventional BOR method even in the interval estimation when there are no outliers.
As the outlier ratio increases, the CPs of our robust methods remain around the nominal level, while those of the BOR method become smaller than the nominal level.
The c-BOR method with the cauchit link has quite low CPs due to the misspecification of the link function, as in the case of log-MSE.

\begin{table}[H]
\caption{Coverage probability of $95\%$ credible intervals for coefficient parameters $\bm{\beta}$}
\begin{center}
\begin{tabular}{cl rr rr c rr c rr}
\hline
& \multicolumn{1}{c}{} & \multicolumn{1}{c}{BOR} & \multicolumn{1}{c}{c-BOR} & \multicolumn{2}{c}{RBOR DDP} & & \multicolumn{2}{c}{RBOR GSP} & & \multicolumn{2}{c}{RBOR GGP} \\ \cline{5-6} \cline{8-9} \cline{11-12}
link&\multicolumn{1}{c}{$\rho$}&&&\multicolumn{1}{c}{0.3}&\multicolumn{1}{c}{0.5}&&\multicolumn{1}{c}{0.3}&\multicolumn{1}{c}{0.5}&&\multicolumn{1}{c}{0.3}&\multicolumn{1}{c}{0.5}\tabularnewline
\hline
&0\%&93.7 & 29.3 & 91.7 & 94.0 && 92.0 & 92.7 && 92.0 & 93.3 \tabularnewline
&5\%&48.3 & 31.2 & 93.3 & 94.2 && 93.5 & 93.5 && 93.5 & 93.7 \tabularnewline
probit&10\%&33.3 & 31.1 & 92.2 & 93.0 && 92.1 & 92.4 && $92.2$&$92.6$ \tabularnewline
&15\%&25.5 & 31.5 & 92.8 & 93.0 && 92.7 & 92.3 && $92.8$&$92.6$ \tabularnewline
&20\%&20.7 & 31.5 & 93.0 & 93.1 && 92.9 & 92.4 && $93.0$&$92.7$ \tabularnewline
\multicolumn{12}{c}{}\tabularnewline
&0\%&$95.3$&$91.0$&$95.0$&$93.3$&&$94.0$&$93.7$&&$94.7$&$94.3$\tabularnewline
&5\%&$62.3$&$67.7$&$93.7$&$92.7$&&$93.0$&$93.5$&&$93.7$&$94.0$\tabularnewline
logit&10\%&$50.6$&$53.7$&$92.8$&$92.3$&&$92.3$&$92.8$&&$92.8$&$93.3$\tabularnewline
&15\%&$45.0$&$47.3$&$92.8$&$92.6$&&$92.4$&$92.9$&&$92.8$&$93.5$\tabularnewline
&20\%&$41.7$&$43.1$&$92.9$&$92.9$&&$92.7$&$93.1$&&$92.9$&$93.5$\tabularnewline
\multicolumn{9}{c}{}\tabularnewline
&0\%&$92.7$&$44.7$&$91.7$&$91.0$&&$91.3$&$90.7$&&$91.3$&$90.7$\tabularnewline
&5\%&$59.2$&$40.7$&$92.3$&$92.2$&&$92.2$&$92.2$&&$92.3$&$92.2$\tabularnewline
loglog&10\%&$48.0$&$35.7$&$92.9$&$92.7$&&$92.7$&$92.4$&&$93.1$&$92.6$\tabularnewline
&15\%&$42.9$&$32.7$&$92.7$&$92.7$&&$92.4$&$92.3$&&$92.8$&$92.3$\tabularnewline
&20\%&$39.8$&$31.0$&$91.9$&$92.1$&&$91.7$&$91.8$&&$92.0$&$91.9$\tabularnewline
\hline
\end{tabular}
\end{center}
\label{tb:cp_beta}
\end{table}%

\begin{table}[H]
\caption{Coverage probability of $95\%$ credible intervals for cutoff parameters $\bm{\delta}$}
\begin{center}
\begin{tabular}{cl rr rr c rr c rr}
\hline
& \multicolumn{1}{c}{} & \multicolumn{1}{c}{BOR} & \multicolumn{1}{c}{c-BOR} & \multicolumn{2}{c}{RBOR DDP} & & \multicolumn{2}{c}{RBOR GSP} & & \multicolumn{2}{c}{RBOR GGP} \\ \cline{5-6} \cline{8-9} \cline{11-12}
link&\multicolumn{1}{c}{$\rho$}&&&\multicolumn{1}{c}{0.3}&\multicolumn{1}{c}{0.5}&&\multicolumn{1}{c}{0.3}&\multicolumn{1}{c}{0.5}&&\multicolumn{1}{c}{0.3}&\multicolumn{1}{c}{0.5}\tabularnewline
\hline
&0\%&$91.0$&$ 5.8$&$90.5$&$91.8$&&$90.0$&$91.8$&&$90.5$&$91.8$\tabularnewline
&5\%&$45.9$&$26.5$&$92.5$&$93.2$&&$92.1$&$93.0$&&$92.6$&$93.1$\tabularnewline
probit&10\%&$30.8$&$41.5$&$91.2$&$91.9$&&$90.8$&$91.7$&&$91.2$&$91.9$\tabularnewline
&15\%&$23.1$&$49.3$&$91.9$&$92.4$&&$91.5$&$92.2$&&$92.0$&$92.5$\tabularnewline
&20\%&$19.0$&$53.9$&$91.4$&$92.0$&&$91.1$&$91.6$&&$91.5$&$91.9$\tabularnewline
\multicolumn{12}{c}{}\tabularnewline
&0\%&$93.8$&$88.0$&$93.2$&$93.2$&&$93.0$&$93.8$&&$93.5$&$94.0$\tabularnewline
&5\%&$55.5$&$74.5$&$92.9$&$93.4$&&$92.5$&$93.4$&&$93.0$&$93.5$\tabularnewline
logit&10\%&$42.3$&$66.1$&$92.8$&$93.4$&&$92.6$&$93.2$&&$92.9$&$93.5$\tabularnewline
&15\%&$35.5$&$61.5$&$92.0$&$92.6$&&$91.8$&$92.1$&&$92.1$&$92.8$\tabularnewline
&20\%&$31.8$&$59.0$&$92.2$&$92.8$&&$92.0$&$92.4$&&$92.2$&$92.9$\tabularnewline
\multicolumn{9}{c}{}\tabularnewline
&0\%&$89.0$&$28.0$&$90.2$&$90.5$&&$90.2$&$90.0$&&$90.2$&$90.5$\tabularnewline
&5\%&$48.9$&$37.9$&$90.6$&$90.8$&&$90.6$&$90.1$&&$90.8$&$90.6$\tabularnewline
loglog&10\%&$36.0$&$40.8$&$91.1$&$91.2$&&$91.0$&$90.8$&&$91.2$&$91.3$\tabularnewline
&15\%&$29.6$&$42.1$&$91.2$&$91.4$&&$91.0$&$90.9$&&$91.3$&$91.3$\tabularnewline
&20\%&$25.8$&$43.3$&$90.9$&$91.0$&&$90.8$&$90.4$&&$91.0$&$90.8$\tabularnewline
\hline
\end{tabular}
\end{center}
\label{tb:cp_delta}
\end{table}%

\begin{rem}[Selection of tuning parameters, $\alpha$ and $\gamma$]
Robust divergences such as the density-power and $\gamma$-divergences have one tuning parameter to control the robustness against outliers (e.g., $\alpha$ and $\gamma$).
If the value of the tuning parameter is smaller than required, the inference result may still be affected by outliers, and conversely, if it is unnecessarily large, the inference may lose the statistical efficiency (\citealp{basu1998robust}).
Therefore, it is necessary to select the optimal tuning parameter, but there are few studies discussing the tuning parameter selection method in the Bayesian inference with the robust divergence.
One method is to use the asymptotic relative efficiency, as in the frequentist framework.
\cite{ghosh2016robust} discuss the case where there is only one parameter to be estimated, and \cite{nakagawa2020default} generalizes the method.
\cite{nakagawa2020default} further state that it is difficult to ensure both the robustness and statistical efficiency when the data are heavily contaminated.
\cite{warwick2005choosing} and \cite{basak2021optimal} propose frequentist methods to chose the optimal value of the tuning parameter using the asymptotic mean squared errors.
Although it may be considered natural to use the model evidence or marginal likelihood as a method to find the optimal tuning parameter in the Bayesian framework, \cite{yonekura2023adaptation} show that such a method is not useful for unnormalized statistical models such as our proposed model.
Recently, \cite{yonekura2023adaptation} proposed a sequential Monte Carlo method (\citealp{del2006sequential}) for simultaneously sampling from the posterior distribution and selecting the tuning parameter in unnormalized statistical models using the Hyv{\"a}rinen score (\citealp{hyvarinen2005estimation}).
They avoid the pilot plot (estimate) problem used in the methods of \cite{warwick2005choosing} and \cite{basak2021optimal} and ensure the stable inference and statistical efficiency.
Applying the method of \cite{yonekura2023adaptation} to the Bayesian ordinal response model requires the development of another computational algorithm based on the sequential Monte Carlo method, and will be the subject of future research.
\end{rem}

\subsection{Real data analysis}
\label{sec:ne_real}
\begin{figure}[H]
\begin{center}
\includegraphics[width=\columnwidth]{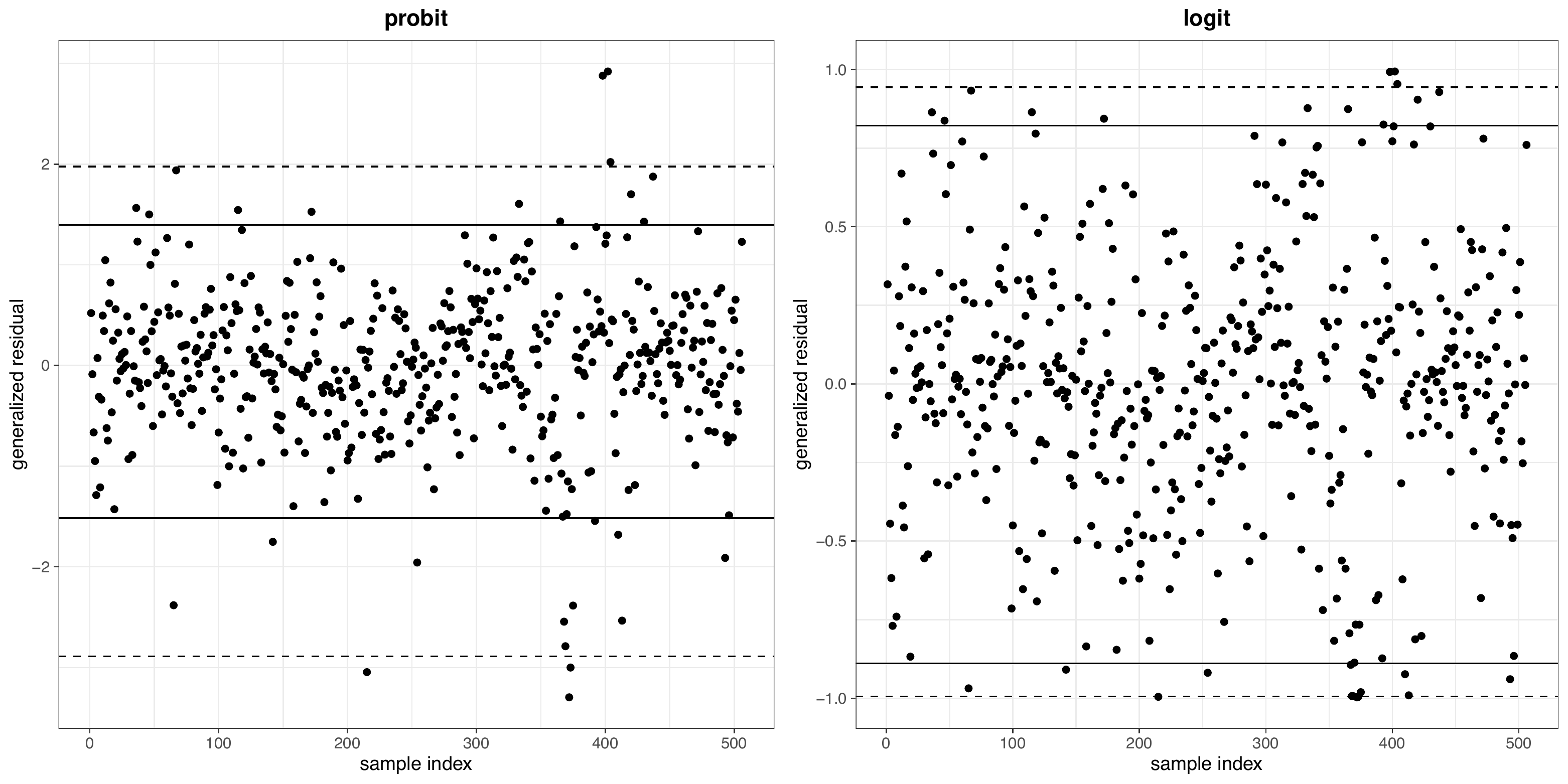}
\caption{The generalized residuals (\citealp{franses2001quantitative}) plots for the Boston housing data with the probit and logit links.
The solid and dashed lines indicate 95\% and 99\% intervals of the generalized residuals, respectively.}
\label{fg:grd_boston}
\end{center}
\end{figure}

\begin{figure}[H]
\begin{center}
\includegraphics[width=\columnwidth]{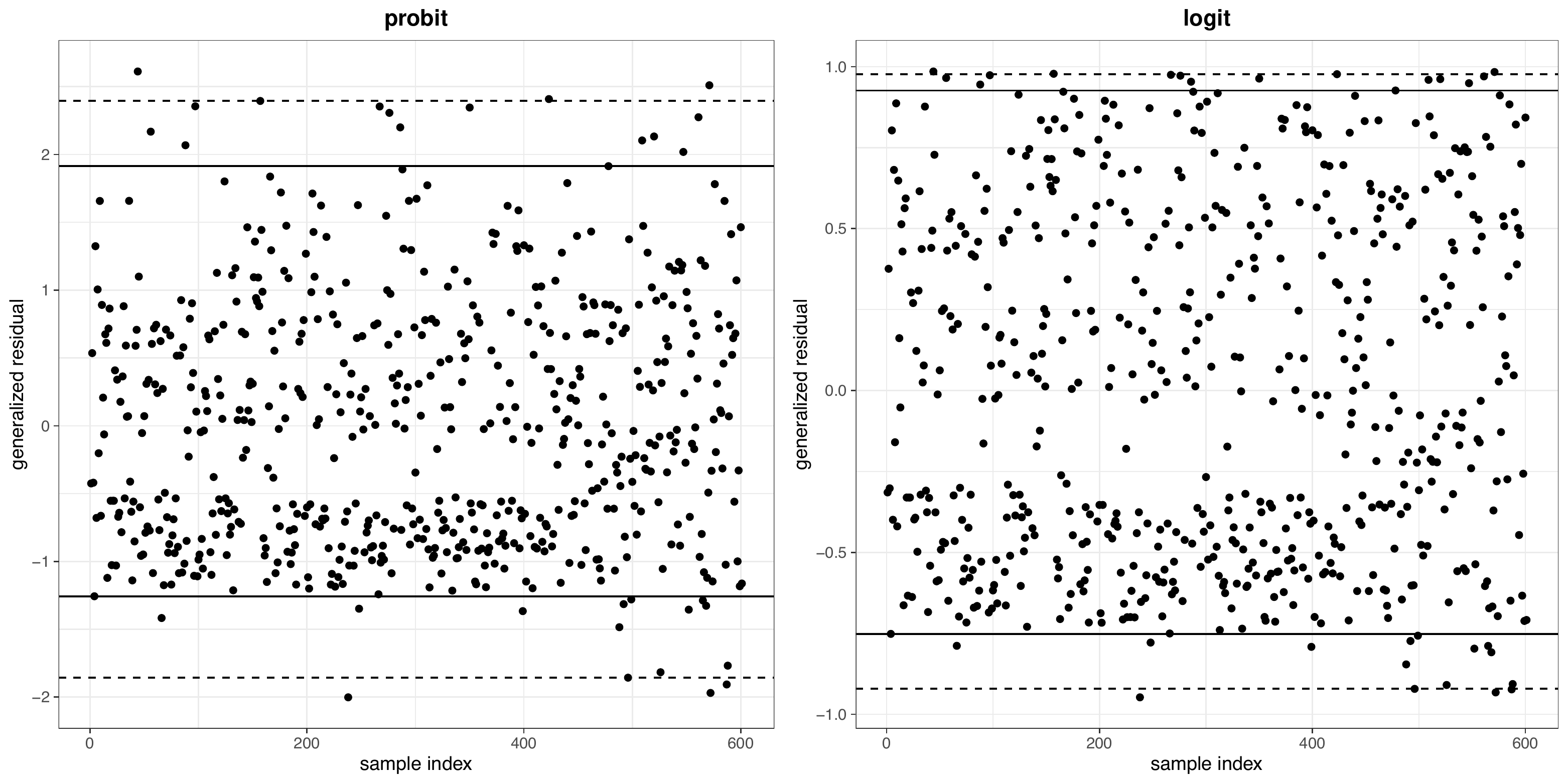}
\caption{The generalized residuals (\citealp{franses2001quantitative}) plots for the Affairs data with the probit and logit links.
Solid and dashed lines indicate 95\% and 99\% intervals of the generalized residuals, respectively.}
\label{fg:grd_affairs}
\end{center}
\end{figure}

This section compares analysis results of our proposed robust Bayesian ordinal response models with those of the non-robust Bayesian ordinal response model through two real datasets: Boston housing data (\citealp{harrison1978hedonic}) and Affairs data (\citealp{fair1978theory}).
The Boston housing data is often used to evaluate the robust inference methods with continuous data as the response variable (\citealp{hashimoto2020robust, hamura2022log, kawakami2023approximate}).
Here, the corrected median value of owner-occupied homes in USD 1000's is transformed into the ordered categorical data with five categories ($\leq10, 10< x \leq 20, 20< x \leq 30, 30< x \leq 40, 40<$), and used it as the response variable.
The Boston housing data has 506 observations and 14 variables, including one binary and 13 continuous variables.
The Affairs data has 601 observations and nine variables, including three continuous, two binary, two ordered categorical, and two multinomial categorical variables.
The self-rating of marriage, one of the ordered categorical data, is set as the response variable.
In the two real data, the continuous data are standardized to have a mean of zero and variance of one, the ordered categorical data in the covariates is numerically transformed using the Likert sigma method, and the multinomial categorical data are transformed using the dummy variables.
In analyzing the two sets of real data, we use the commonly used symmetric link functions, the probit and logit links.
Figures \ref{fg:grd_boston} and \ref{fg:grd_affairs} show the generalized residuals (\citealp{franses2001quantitative}) plots with two link functions for two real datasets.
The 95\% and 99\% intervals of the generalized residuals are indicated in these figures as the solid and dashed lines, respectively.
As can be seen from these figures, since some of the data are well outside the solid line of the 95\% interval, the Boston housing data and the Affairs data may contain outliers.

\begin{figure}[H]
  \begin{minipage}[b]{0.5\linewidth}
    \centering
    \includegraphics[keepaspectratio, width=\columnwidth]{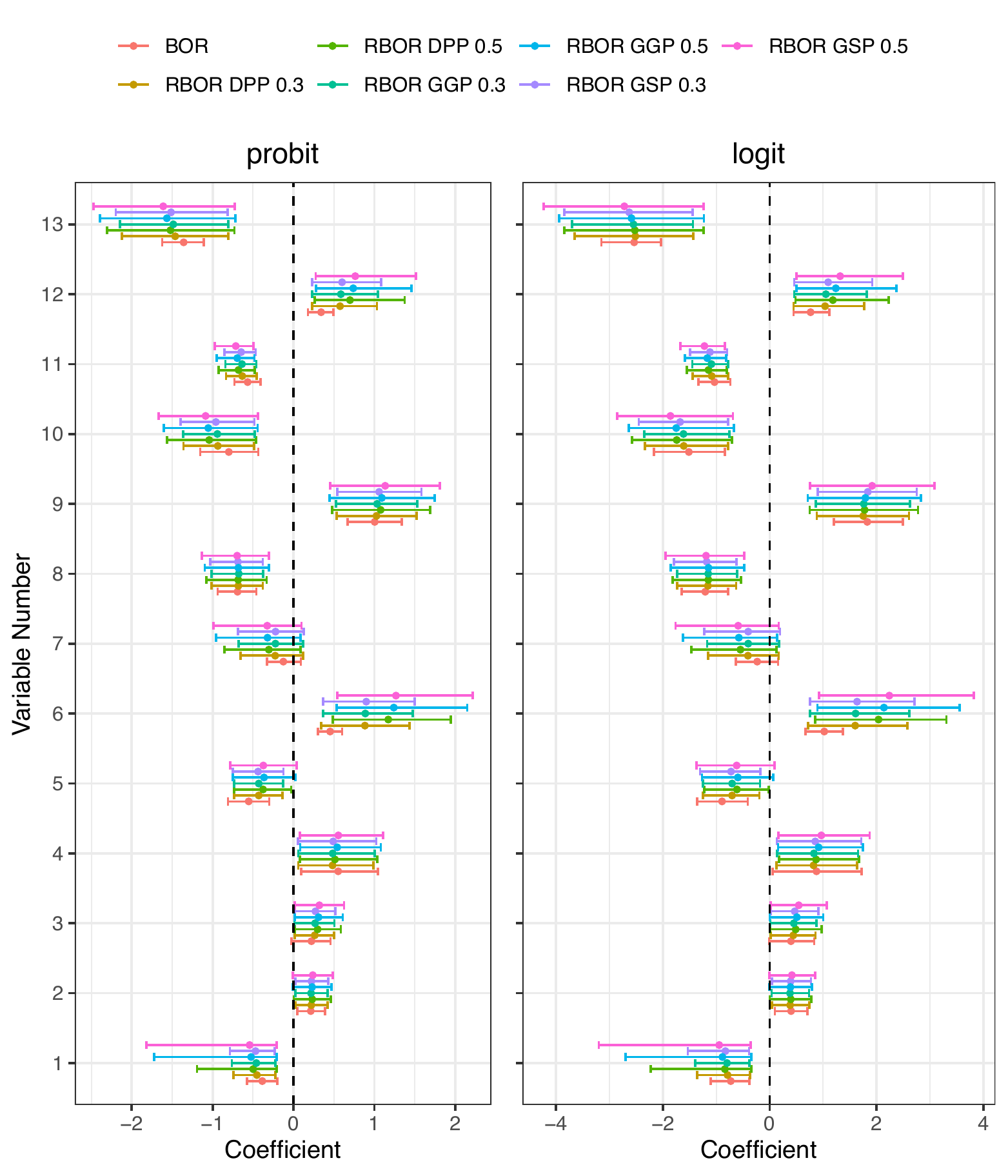}
    \subcaption{Boston housing data}
  \end{minipage}
  \begin{minipage}[b]{0.5\linewidth}
    \centering
    \includegraphics[keepaspectratio, width=\columnwidth]{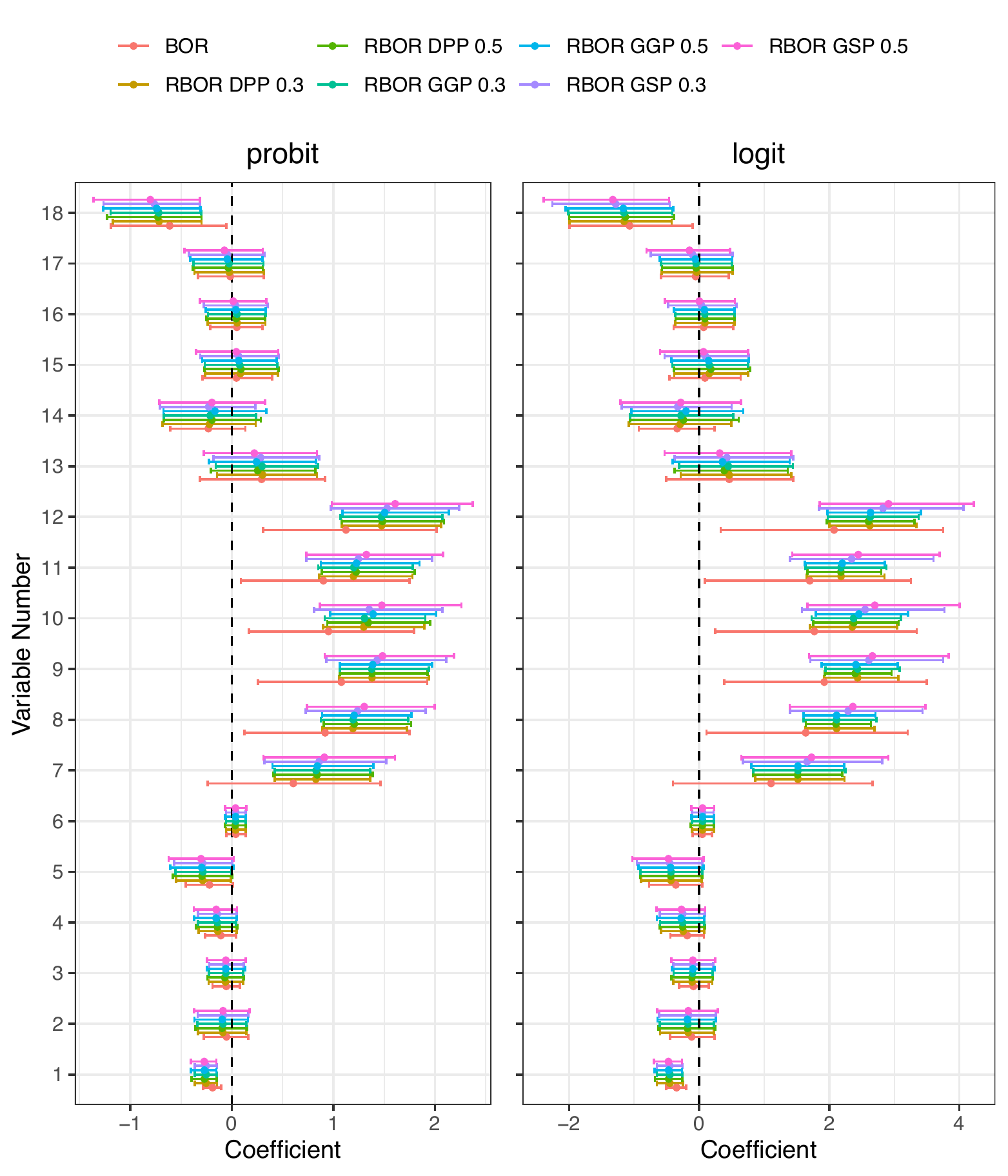}
    \subcaption{Affairs data}
  \end{minipage}
  \caption{
  95\% credible intervals and posterior medians of the coefficient parameters applying the proposed robust Bayesian ordinal response (RBOR DPP, GSP, and GGP) model and the non-robust Bayesian ordinal response (BOR) model with the probit and logit link to the Boston housing data and the Affairs data.
  }
  \label{fg:da_rborm_coef}
\end{figure}

\begin{figure}[H]
  \begin{minipage}[b]{0.5\linewidth}
    \centering
    \includegraphics[keepaspectratio, width=\columnwidth]{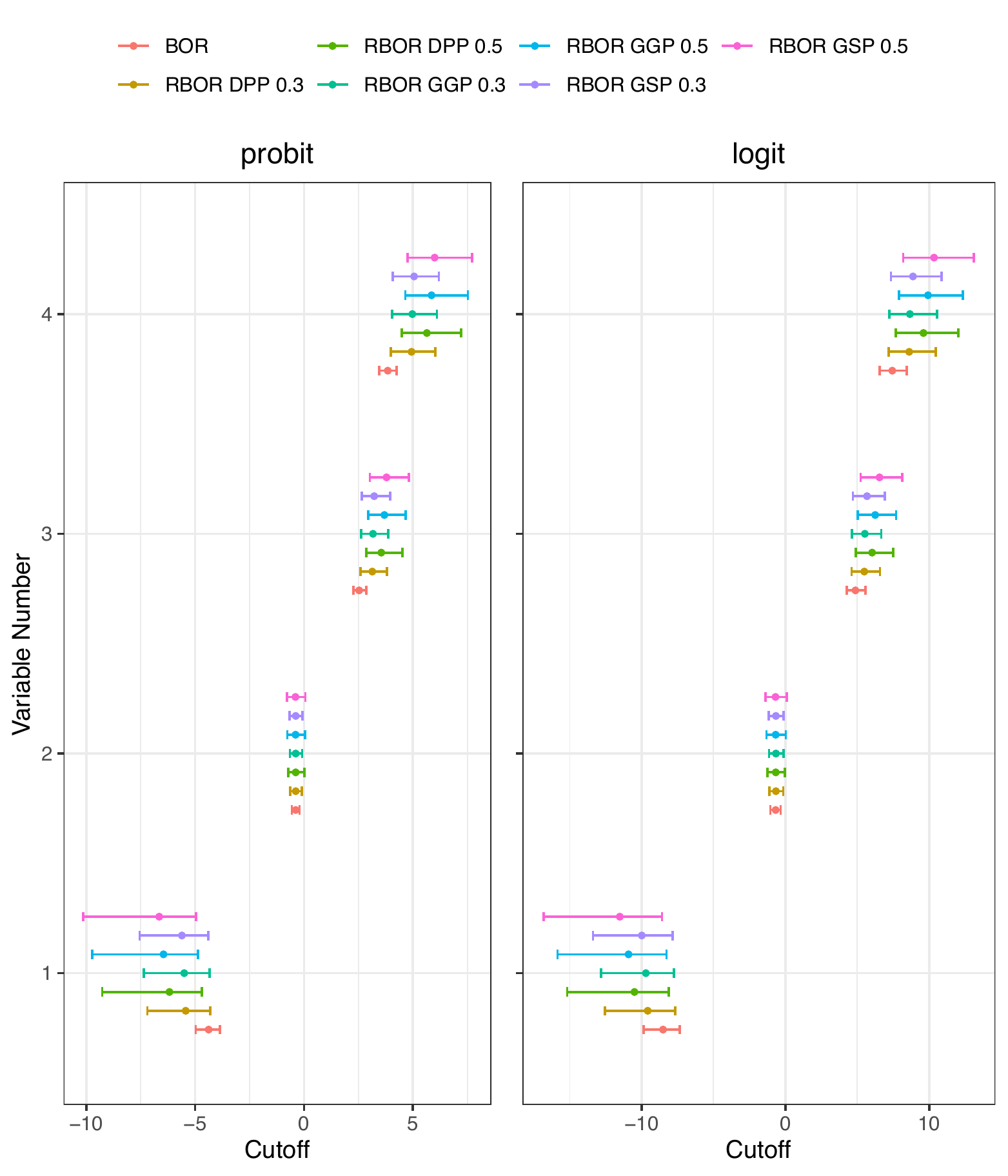}
    \subcaption{Boston housing data}
  \end{minipage}
  \begin{minipage}[b]{0.5\linewidth}
    \centering
    \includegraphics[keepaspectratio, width=\columnwidth]{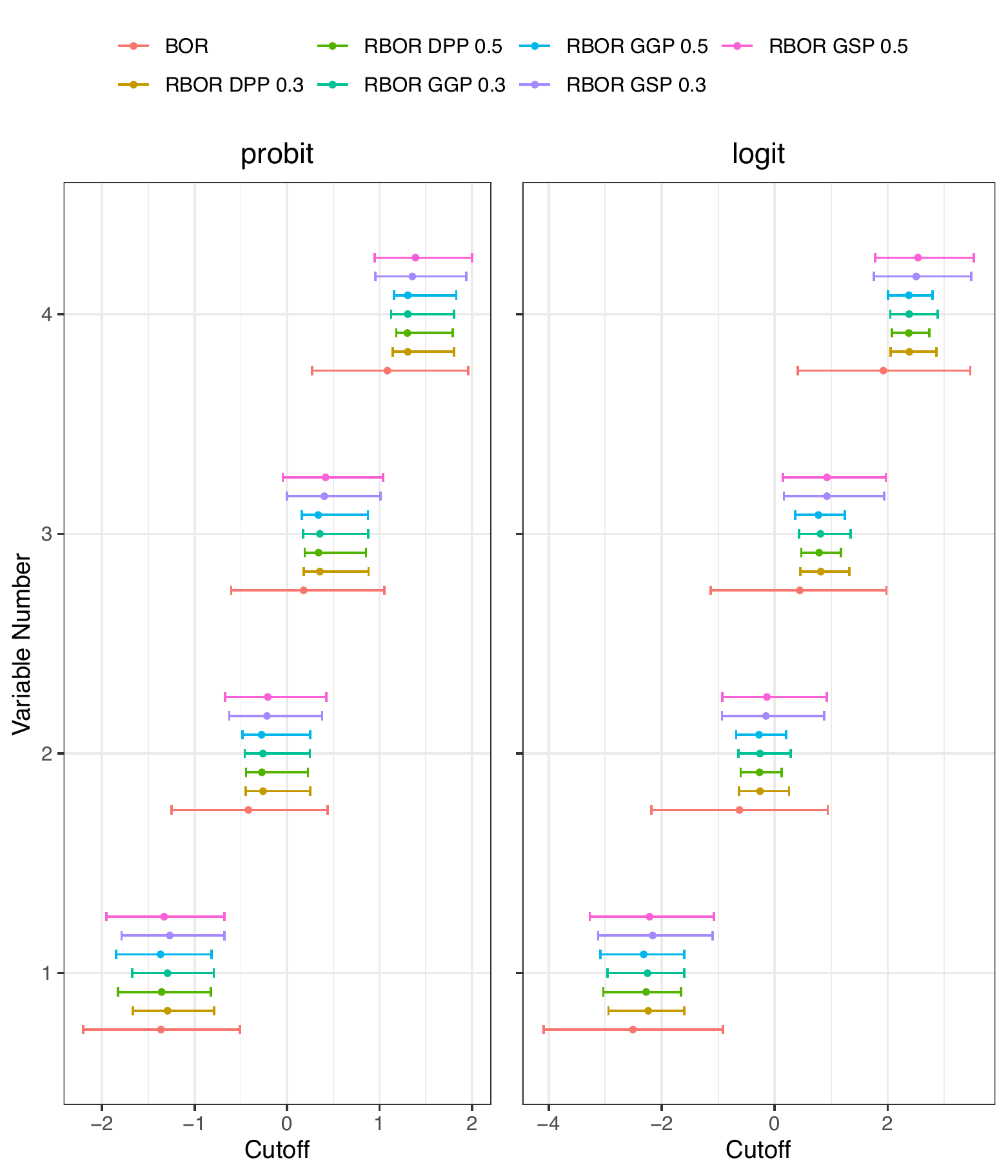}
    \subcaption{Affairs data}
  \end{minipage}
  \caption{
  95\% credible intervals and posterior medians of the cutoff parameters applying the proposed robust Bayesian ordinal response (RBOR DPP, GSP, and GGP) model and the non-robust Bayesian ordinal response (BOR) model with the probit and logit link to the Boston housing data and the Affairs data.
  }
  \label{fg:da_rborm_cut}
\end{figure}

We apply our proposed Bayesian ordinal robust model with the density-power posterior (RBOR DPP), $\gamma$-synthetic posterior (GSP), and $\gamma$-general posterior (GGP) to these data and generate 2000 posterior samples using the WLB method.
The tuning parameters ($\alpha$ and $\gamma$) for the density-power and $\gamma$-divergences are set to 0.3 and 0.5.
For comparison, we apply the non-robust Bayesian ordinal response model with the standard posterior (BOR) and generate 2500 posterior samples using the function \texttt{stan\_polr} of the {\bf rstanarm} package in the {\bf R}, 500 of which are burn-in.
The posterior medians and 95\% credible intervals are calculated using the 2000 posterior samples obtained for all methods.
Figures \ref{fg:da_rborm_coef} and \ref{fg:da_rborm_cut} show the 95\% credible intervals and posterior medians of the coefficient and cutoff parameters applying the proposed robust Bayesian ordinal response model and the non-robust Bayesian ordinal response model with the probit and logit link to the Boston housing data and the Affairs data.
From these figures, we can see that the proposed robust method, RBOR, and the non-robust method, BOR, show different results in the inference of some parameters.
Particularly, in Figure \ref{fg:da_rborm_coef}, for variable number 3 in the Boston housing data and variable number 7 in the Affairs data, the credible intervals in the BOR includes 0, but not in the RBOR.
Additionally, for variables number 2 and 5 in the Boston housing data, the credible intervals in the BOR do not include 0, but some of the RBORs do.
Since the positions of the credible intervals and the posterior medians of the parameters are different in many cases between the BOR and RBOR, both the Boston housing data and the Affairs data may contain outliers.

\section{Conclusion and remarks}
\label{sec:conclude}
This article addressed the problem induced by outliers on the ordinal response model, which is a regression on an ordered categorical data as the response variable, in the Bayesian framework.
Through the application of the Affairs data (\citealp{fair1978theory}) (Example \ref{exm:post} in Section \ref{sec:borm}), we confirmed that the posterior distribution in the conventional Bayesian ordinal response model based on the likelihood \eqref{eq:post} does not satisfy the posterior robustness, i.e., the posterior distribution differs between cases with and without large outliers.
Furthermore, we proved that the standard posterior cannot satisfy the posterior robustness with the distribution function (link function) of any distribution that belongs to the parametric distribution (Theorem \ref{thm:stand_postrobust} in Section \ref{sec:stand_postrobust}).
Meanwhile, the robustness against outliers on the ordinal response model in the framework of the frequentist has been addressed by \cite{scalera2021robust} and \cite{momozaki2022robustness}.
\cite{scalera2021robust} derives conditions for the link function such that the influence function satisfies the boundedness, and shows that the Student's $t$ and Cauchy distributions satisfy the conditions.
\cite{momozaki2022robustness} proved that the influence function cannot satisfy the redescendence, i.e., it cannot completely eliminate the influence of large outliers in the ordinal response model with arbitrary link function.
Thus, the redescendence of the influence function in the frequentist framework and the posterior robustness in the Bayesian framework may be related.

In order to achieve robust Bayesian inference in the ordinal response model, we proposed state-of-the-art robust Bayesian ordinal response models with two robust divergences, the density-power and $\gamma$-divergences, based on the general (synthetic) posterior inference.
We also proved that our proposed posteriors satisfy the posterior robustness (Theorems \ref{thm:dp_postrobust}, \ref{thm:gamma_HS_postrobust}, and \ref{thm:gamma_NH_postrobust} in Section \ref{sec:post_robust}).
Furthermore, we proposed an algorithm for generating posterior samples from our proposed posteriors using the weighted likelihood bootstrap (WLB) method.
As described in the remark in Section \ref{sec:post_comp}, it is possible to generate the posterior samples using the probabilistic programming language {\bf Stan} based on the Hamiltonian Monte Carlo method instead of the WLB method.
However, we recommend the use of the algorithm based on the WLB method because the algorithm based on the WLB method generates independent samples and thus can ignore the mixing problem in the posterior samples, and the algorithm based on the {\bf Stan} does not work well when the dimension of parameters is large.
Using the index of robustness based on the Fisher-Rao metric proposed in \cite{kurtek2015bayesian}, we also showed that our proposed robust methods are robust against outliers, i.e., can ignore the influence of large outliers (Section \ref{sec:ne_robust}).
Furthermore, using several artificial data with varying outlier ratios, we demonstrated that our proposed methods perform as well as the conventional method when there are no outliers, and that as the outlier ratio increases, the conventional method performs worse, while our proposed methods still perform well (Section \ref{sec:ne_simu}).
In applications using two real datasets, the Boston housing and Affairs data, that seem to include outliers from the values of the generalized residuals, our proposed methods and the conventional method yield different analysis results in terms of significant variables (Section \ref{sec:ne_real}).

The posteriors in our proposed robust Bayesian ordinal response model satisfy the posterior robustness even with the commonly used link functions such as the probit, logit, and complementary loglog links.
This fact is a significant contribution to the analysts using the ordinal response model as it provides them with robust and flexible analysis methods for data with outliers.
As confirmed in the numerical experiments in Section \ref{sec:ne_simu}, the misspecification of the link function causes a substantial bias in the inference.
It would be very important for analysts to be able to choose a link function that suits their field of study.
Furthermore, our study will provide a basis for developing an inference method that is robust to both the misspecification of the link function and outliers.
This is also our future work.

Finally, the choice of prior distribution is an important topic in the Bayesian inference.
When there is no prior information on the parameters in a model, we may consider an objective prior distribution.
\cite{giummole2019objective} and \cite{leisen2020class} derived objective prior distributions based on scoring rules.
\cite{nakagawa2020default} derived the reference prior and the moment matching prior in the inference with the $\gamma$-divergence.
The derivation of the objective prior in our proposed robust Bayesian ordinal response model is a subject for future work.

The {\bf R} code implementing the proposed robust Bayesian method for the ordinal response model is available at GitHub repository (\url{https://github.com/t-momozaki/RBORM}).

%=acknowledgement=========================================================
\section*{Acknowledgement}
This work was JSPS Grant-in-Aid for Early-Career Scientists Grant Number JP19K14597, JSPS Grant-in-Aid for Scientific Research (C) Number JP20K03756, JSPS Grant-in-Aid for Scientific Research (B) Number 21H00699, and JSPS Grant-in-Aid for Early-Career Scientists Grant Number JP23K13019.

%=\appendix==========================================
\def\thesection{Appendix}
\def\thesubsection{A.\arabic{subsection}}
\section{}
\setcounter{equation}{0}
\def\theequation{A.\arabic{equation}}%%数式の番号付け(section, 式番号)
\def\thethm{A.\arabic{thm}}%%数式の番号付け(section, 式番号)
\def\thelem{A.\arabic{lem}}%%数式の番号付け(section, 式番号)

\subsection{Proof of the theorems in Section \ref{sec:post_robust}}
\label{app:post_robust}

\subsubsection{Proof of Theorem \ref{thm:stand_postrobust}}
\label{app:stand_postrobust}
Consider
\begin{equation*}
\frac{ p(\bm{\theta}|D) }{ p(\bm{\theta}|D^*) } = \frac{ p(D^*) }{ p(D) } \frac{ \prod_{i=1}^n f(y_i|\bm{x}_i;\bm{\theta}) }{ \prod_{i\in\mathcal{K}} f(y_i|\bm{x}_i;\bm{\theta}) },
\end{equation*}
where $p(D) = \int p(\bm{\theta}) \prod_{i=1}^n f(y_i|\bm{x}_i;\bm{\theta}) d\bm{\theta}$ and $p(D^*) = \int p(\bm{\theta}) \prod_{i\in\mathcal{K}} f(y_i|\bm{x}_i;\bm{\theta}) d\bm{\theta}$.

Since from the Lagrange's theorem $G(\delta_{y_i}-\bm{x}_i^\top\bm{\beta})-G(\delta_{y_i-1}-\bm{x}_i^\top\bm{\beta}) = (\delta_{y_i}-\delta_{y_i-1}) g(\delta_{i}^*-\bm{x}_i^\top\bm{\beta})$ for $\delta_{i}^*$ such that $\delta_{y_i-1}<\delta_{i}^*<\delta_{y_i}$,
\begin{align*}
\frac{ p(\bm{\theta}|D) }{ p(\bm{\theta}|D^*) } &= \frac{ p(D^*) }{ p(D) } \prod_{i\in\mathcal{L}} (\delta_{y_i}-\delta_{y_i-1}) g(\delta_{i}^*-\bm{x}_i^\top\bm{\beta}) \\
&= \left[ \frac{ p(D^*) }{ p(D) } \prod_{i\in\mathcal{L}} g(d_i) \right] \left[ \prod_{i\in\mathcal{L}} \frac{ (\delta_{y_i}-\delta_{y_i-1}) g(\delta_{i}^*-\bm{x}_i^\top\bm{\beta}) }{ g(d_i) } \right], 
\end{align*}
where $d_i$ is independent variable of the parameter $\bm{\theta}$, and 
\begin{align*}
&\frac{ p(D) }{ p(D^*) \prod_{i\in\mathcal{L}} g(d_i) } \\
&= \frac{ p(D) }{ p(D^*) \prod_{i\in\mathcal{L}} g(d_i) } \int p(\bm{\theta} | D) d\bm{\theta} \\
&= \int \frac{ p(\bm{\theta}) \prod_{i\in\mathcal{K}} (\delta_{y_i}-\delta_{y_i-1}) g(\delta_{i}^*-\bm{x}_i^\top\bm{\beta}) }{ p(D^*) \prod_{i\in\mathcal{L}} g(d_i) } \left[ \prod_{i\in\mathcal{L}} (\delta_{y_i}-\delta_{y_i-1}) g(\delta_{i}^*-\bm{x}_i^\top\bm{\beta}) \right] d\bm{\theta} \\
&= \int p(\bm{\theta}|D^*) \prod_{i\in\mathcal{L}} \frac{ (\delta_{y_i}-\delta_{y_i-1}) g(\delta_{i}^*-\bm{x}_i^\top\bm{\beta}) }{ g(d_i) } d\bm{\theta}.
\end{align*}

Therefore, for the posterior robustness to hold, it must be 
\begin{equation*}
\frac{ (\delta_{y_i}-\delta_{y_i-1}) g(\delta_{i}^*-\bm{x}_i^\top\bm{\beta}) }{ g(d_i) } = O(1), 
\end{equation*}
which requires that $g(\cdot)$ is a regularly varying function (\citealp{o2012bayesian}).
However, for $g(\cdot)$ to be a regularly varying function, its tail order must be smaller than $1/x$, and such a distribution does not exist.
Namely, no matter what distribution $G(\cdot)$ is used, the posterior robustness does not hold for the standard posterior in the ordinal response model.

\qed

\subsubsection{Proof of Theorem \ref{thm:dp_postrobust}}
\label{app:dp_postrobust}
Consider
\begin{equation*}
\frac{ p_{DP}(\bm{\theta}|D) }{ p_{DP}(\bm{\theta}|D^*) } = \frac{ p_{DP}(D^*) }{ p_{DP}(D) } \frac{ \exp\left\{ R_{DP}(D|\bm{\theta}) \right\} }{ \exp\left\{ R_{DP}(D^*|\bm{\theta}) \right\} },
\end{equation*}
where $p_{DP}(D^*) = \int p(\bm{\theta}) \exp\left\{ R_{DP}(D^*|\bm{\theta}) \right\} d\bm{\theta}$ and $R_{DP}(D^*|\bm{\theta}) = \sum_{i\in\mathcal{K}} r_{DP}(y_i|\bm{x}_i;\bm{\theta})$.

Since $\lim_{\omega\to\infty} r_{DP}(y_i|\bm{x}_i;\bm{\theta}) = -(1+\alpha)^{-1}$, then
\begin{align*}
&\lim_{\omega\to\infty} R_{DP}(D|\bm{\theta}) \\
=& \lim_{\omega\to\infty} \sum_{i=1}^n r_{DP}(y_i|\bm{x}_i;\bm{\theta}) \\
=& - \frac{|\mathcal{L}|}{1+\alpha} + \\
&\sum_{i\in\mathcal{K}} \left\{ \frac{1}{\alpha} \left[ G(\delta_{y_i}-\bm{x}_i^\top\bm{\beta}) - G(\delta_{y_i-1}-\bm{x}_i^\top\bm{\beta}) \right]^\alpha - \frac{1}{1+\alpha} \sum_{m=1}^M \left[ G(\delta_{m}-\bm{x}_i^\top\bm{\beta}) - G(\delta_{m-1}-\bm{x}_i^\top\bm{\beta}) \right]^{1+\alpha} \right\} \\
=& - \frac{|\mathcal{L}|}{1+\alpha} + \sum_{i\in\mathcal{K}} \exp\left\{ r_{DP}(y_i|\bm{x}_i;\bm{\theta}) \right\} = - \frac{|\mathcal{L}|}{1+\alpha} + R_{\gamma(2)}(D^*|\bm{\theta}),
\end{align*}
where $|\mathcal{L}|$ is the number of outliers and
\begin{align*}
\lim_{\omega\to\infty} \frac{ \exp\left\{ R_{\gamma(2)}(D|\bm{\theta}) \right\} }{ \exp\left\{ R_{\gamma(2)}(D^*|\bm{\theta}) \right\} } = \exp\left\{ - \frac{|\mathcal{L}|}{1+\alpha} \right\}.
\end{align*}

Next, consider
\begin{equation*}
\lim_{\omega\to\infty} p_{DP}(D) = \lim_{\omega\to\infty} \int p(\bm{\theta}) \exp\left\{ R_{DP}(D|\bm{\theta}) \right\} d\bm{\theta}.
\end{equation*}
Since $G(\delta_{y_i}-\bm{x}_i^\top\bm{\beta}) - G(\delta_{y_i-1}-\bm{x}_i^\top\bm{\beta})$ is bounded, where $G(\cdot)$ is the distribution function,
\begin{align*}
&R_{DP}(D|\bm{\theta}) \\
&= \sum_{i=1}^n \left\{ \frac{1}{\alpha} \left[ G(\delta_{y_i}-\bm{x}_i^\top\bm{\beta}) - G(\delta_{y_i-1}-\bm{x}_i^\top\bm{\beta}) \right]^\alpha - \frac{1}{1+\alpha} \sum_{m=1}^M \left[ G(\delta_{m}-\bm{x}_i^\top\bm{\beta}) - G(\delta_{m-1}-\bm{x}_i^\top\bm{\beta}) \right]^{1+\alpha} \right\}
\end{align*}
is also clearly bounded.
Hence since $p(\bm{\theta}) \exp\left\{ R_{DP}(D|\bm{\theta}) \right\} < \infty$ under the proper prior $p(\bm{\theta})$, from Lebesgue's dominated convergence theorem, 
\begin{align*}
\lim_{\omega\to\infty} p_{DP}(D) &= \lim_{\omega\to\infty} \int p(\bm{\theta}) \exp\left\{ R_{DP}(D|\bm{\theta}) \right\} d\bm{\theta} \\
&= \int p(\bm{\theta}) \left( \lim_{\omega\to\infty} \exp\left\{ R_{DP}(D|\bm{\theta}) \right\} \right) d\bm{\theta} \\
&= \int p(\bm{\theta}) \exp\left\{ - \frac{|\mathcal{L}|}{1+\alpha} \right\} \exp\left\{ R_{DP}(D^*|\bm{\theta}) \right\} d\bm{\theta} = \exp\left\{ - \frac{|\mathcal{L}|}{1+\alpha} \right\} p_{\gamma(2)}(D^*),
\end{align*}
and then
\begin{equation*}
\lim_{\omega\to\infty} \frac{ p_{DP}(\bm{\theta}|D) }{ p_{DP}(\bm{\theta}|D^*) } = \exp\left\{  \frac{|\mathcal{L}|}{1+\alpha} \right\} \exp\left\{ - \frac{|\mathcal{L}|}{1+\alpha} \right\} = 1,
\end{equation*}
so the posterior robustness holds.

\qed

\subsubsection{Proof of Theorem \ref{thm:gamma_HS_postrobust}}
\label{app:gamma_HS_postrobust}
Consider
\begin{equation*}
\frac{ p_{\gamma(1)}(\bm{\theta}|D) }{ p_{\gamma(1)}(\bm{\theta}|D^*) } = \frac{ p_{\gamma(1)}(D^*) }{ p_{\gamma(1)}(D) } \frac{ \exp\left\{ R_{\gamma(1)}(D|\bm{\theta}) \right\} }{ \exp\left\{ R_{\gamma(1)}(D^*|\bm{\theta}) \right\} },
\end{equation*}
where $p_{\gamma(1)}(D^*) = \int p(\bm{\theta}) \exp\left\{ R_{\gamma(1)}(D^*|\bm{\theta}) \right\} d\bm{\theta}$ and $R_{\gamma(1)}(D^*|\bm{\theta}) = \frac{n}{\gamma} \log\left\{ \frac{\gamma}{n} \sum_{i\in\mathcal{K}} r_{\gamma}(y_i|\bm{x}_i;\bm{\theta}) \right\}$.

Since $\lim_{\omega\to\infty} r_{\gamma}(y_i|\bm{x}_i;\bm{\theta}) = 0$, then
\begin{align*}
\lim_{\omega\to\infty} R_{\gamma(1)}(D|\bm{\theta}) &= \lim_{\omega\to\infty} \frac{n}{\gamma} \log\left\{ \frac{\gamma}{n} \sum_{i=1}^n r_{\gamma}(y_i|\bm{x}_i;\bm{\theta}) \right\} \\
&= \frac{n}{\gamma} \log\left\{ \frac{\gamma}{n} \sum_{i\in\mathcal{K}} r_{\gamma}(y_i|\bm{x}_i;\bm{\theta}) \right\} = R_{\gamma(1)}(D^*|\bm{\theta})
\end{align*}
and
\begin{align*}
\lim_{\omega\to\infty} \frac{ \exp\left\{ R_{\gamma(1)}(D|\bm{\theta}) \right\} }{ \exp\left\{ R_{\gamma(1)}(D^*|\bm{\theta}) \right\} } = 1.
\end{align*}

Next, consider
\begin{equation*}
\lim_{\omega\to\infty} p_{\gamma(1)}(D) = \lim_{\omega\to\infty} \int p(\bm{\theta}) \exp\left\{ R_{\gamma(1)}(D|\bm{\theta}) \right\} d\bm{\theta}.
\end{equation*}
Since $G(\delta_{y_i}-\bm{x}_i^\top\bm{\beta}) - G(\delta_{y_i-1}-\bm{x}_i^\top\bm{\beta})$ is bounded, where $G(\cdot)$ is the distribution function,
\begin{equation*}
R_{\gamma(1)}(D|\bm{\theta}) = \frac{n}{\gamma} \log\left\{ \frac{1}{n} \sum_{i=1}^n \left( \frac{ G(\delta_{y_i}-\bm{x}_i^\top\bm{\beta}) - G(\delta_{y_i-1}-\bm{x}_i^\top\bm{\beta}) }{ \left( \sum_{m=1}^M \left[ G(\delta_{m}-\bm{x}_i^\top\bm{\beta}) - G(\delta_{m-1}-\bm{x}_i^\top\bm{\beta}) \right]^{1+\gamma} \right)^{1/(1+\gamma)} } \right)^{\gamma} \right\}
\end{equation*}
is also clearly bounded.
Hence since $p(\bm{\theta}) \exp\left\{ R_{\gamma(1)}(D|\bm{\theta}) \right\} < \infty$ under the proper prior $p(\bm{\theta})$, from Lebesgue's dominated convergence theorem, 
\begin{align*}
\lim_{\omega\to\infty} p_{\gamma(1)}(D) &= \lim_{\omega\to\infty} \int p(\bm{\theta}) \exp\left\{ R_{\gamma(1)}(D|\bm{\theta}) \right\} d\bm{\theta} \\
&= \int p(\bm{\theta}) \left( \lim_{\omega\to\infty} \exp\left\{ R_{\gamma(1)}(D|\bm{\theta}) \right\} \right) d\bm{\theta} \\
&= \int p(\bm{\theta}) \exp\left\{ R_{\gamma(1)}(D^*|\bm{\theta}) \right\} d\bm{\theta} = p_{\gamma(1)}(D^*),
\end{align*}
and then the posterior robustness
\begin{equation*}
\lim_{\omega\to\infty} p_{\gamma(1)}(\bm{\theta}|D) = p_{\gamma(1)}(\bm{\theta}|D^*)
\end{equation*}
holds.

\qed

\subsubsection{Proof of Theorem \ref{thm:gamma_NH_postrobust}}
\label{app:gamma_NH_postrobust}
Consider
\begin{equation*}
\frac{ p_{\gamma(2)}(\bm{\theta}|D) }{ p_{\gamma(2)}(\bm{\theta}|D^*) } = \frac{ p_{\gamma(2)}(D^*) }{ p_{\gamma(2)}(D) } \frac{ \exp\left\{ R_{\gamma(2)}(D|\bm{\theta}) \right\} }{ \exp\left\{ R_{\gamma(2)}(D^*|\bm{\theta}) \right\} },
\end{equation*}
where $p_{\gamma(2)}(D^*) = \int p(\bm{\theta}) \exp\left\{ R_{\gamma(2)}(D^*|\bm{\theta}) \right\} d\bm{\theta}$ and $R_{\gamma(2)}(D^*|\bm{\theta}) = \prod_{i\in\mathcal{K}} \exp\left\{ r_{\gamma}(y_i|\bm{x}_i;\bm{\theta}) \right\}$.

Since $\lim_{\omega\to\infty} r_{\gamma}(y_i|\bm{x}_i;\bm{\theta}) = 0$, then
\begin{align*}
\lim_{\omega\to\infty} R_{\gamma(2)}(D|\bm{\theta}) &= \lim_{\omega\to\infty} \sum_{i=1}^n r_{\gamma}(y_i|\bm{x}_i;\bm{\theta}) \\
&= \sum_{i\in\mathcal{K}} r_{\gamma}(y_i|\bm{x}_i;\bm{\theta}) = R_{\gamma(2)}(D^*|\bm{\theta})
\end{align*}
and
\begin{align*}
\lim_{\omega\to\infty} \frac{ \exp\left\{ R_{\gamma(2)}(D|\bm{\theta}) \right\} }{ \exp\left\{ R_{\gamma(2)}(D^*|\bm{\theta}) \right\} } = 1.
\end{align*}

Next, consider
\begin{equation*}
\lim_{\omega\to\infty} p_{\gamma(2)}(D) = \lim_{\omega\to\infty} \int p(\bm{\theta}) \exp\left\{ R_{\gamma(2)}(D|\bm{\theta}) \right\} d\bm{\theta}.
\end{equation*}
Since $G(\delta_{y_i}-\bm{x}_i^\top\bm{\beta}) - G(\delta_{y_i-1}-\bm{x}_i^\top\bm{\beta})$ is bounded, where $G(\cdot)$ is the distribution function,
\begin{equation*}
R_{\gamma(2)}(D|\bm{\theta}) = \sum_{i=1}^n \left\{ \frac{1}{\gamma} \left( \frac{ G(\delta_{y_i}-\bm{x}_i^\top\bm{\beta}) - G(\delta_{y_i-1}-\bm{x}_i^\top\bm{\beta}) }{ \left( \sum_{m=1}^M \left[ G(\delta_{m}-\bm{x}_i^\top\bm{\beta}) - G(\delta_{m-1}-\bm{x}_i^\top\bm{\beta}) \right]^{1+\gamma} \right)^{1/(1+\gamma)} } \right)^{\gamma} \right\}
\end{equation*}
is also clearly bounded.
Hence since $p(\bm{\theta}) \exp\left\{ R_{\gamma(2)}(D|\bm{\theta}) \right\} < \infty$ under the proper prior $p(\bm{\theta})$, from Lebesgue's dominated convergence theorem, 
\begin{align*}
\lim_{\omega\to\infty} p_{\gamma(2)}(D) &= \lim_{\omega\to\infty} \int p(\bm{\theta}) \exp\left\{ R_{\gamma(2)}(D|\bm{\theta}) \right\} d\bm{\theta} \\
&= \int p(\bm{\theta}) \left( \lim_{\omega\to\infty} \exp\left\{ R_{\gamma(2)}(D|\bm{\theta}) \right\} \right) d\bm{\theta} \\
&= \int p(\bm{\theta}) \exp\left\{ R_{\gamma(2)}(D^*|\bm{\theta}) \right\} d\bm{\theta} = p_{\gamma(2)}(D^*),
\end{align*}
and then the posterior robustness
\begin{equation*}
\lim_{\omega\to\infty} p_{\gamma(2)}(\bm{\theta}|D) = p_{\gamma(2)}(\bm{\theta}|D^*)
\end{equation*}
holds.

\qed

%=reference=========================================================
\bibliographystyle{myapalike2} 
\bibliography{References.bib}
\end{document}